\documentclass[preprint]{aastex}
\usepackage{epsf}
\usepackage{emulateapj5}
\usepackage{onecolfloat}

\newcommand{\beqa}{\begin{eqnarray}}
\newcommand{\eeqa}{\end{eqnarray}}

\def\gsim {\lower .1ex\hbox{\rlap{\raise .6ex\hbox{\hskip .3ex
        {\ifmmode{\scriptscriptstyle >}\else
                {$\scriptscriptstyle >$}\fi}}}
        \kern -.4ex{\ifmmode{\scriptscriptstyle \sim}\else
                {$\scriptscriptstyle\sim$}\fi}}}
\def\lsim {\lower .1ex\hbox{\rlap{\raise .6ex\hbox{\hskip .3ex
        {\ifmmode{\scriptscriptstyle <}\else
                {$\scriptscriptstyle <$}\fi}}}
        \kern -.4ex{\ifmmode{\scriptscriptstyle \sim}\else
                {$\scriptscriptstyle\sim$}\fi}}}
\def\beq{\begin{equation}}
\def\eeq{\end{equation}}

\begin{document}
\slugcomment{{\em Astrophysical Journal, submitted}}
\twocolumn[

\lefthead{Faint AGN and the Ionizing Background}
\righthead{SCHIRBER and BULLOCK}

\title{Faint AGN and the Ionizing Background}\vspace{3mm}

\author{Michael Schirber and James S. Bullock}
 \affil{Department of Physics,    The  Ohio State
University, 174 W. 18th Ave, Columbus, OH 43210-1173} 

\begin{abstract}

We  determine the   evolution  of the   faint, high-redshift,  optical
luminosity function of Active Galactic Nuclei (AGN) implied by several
observationally-motivated models of the  ionizing background from $3 <
z  < 5$.  Our  results depend crucially  on  whether we  use the total
ionizing rate measured by the proximity effect  technique or the lower
determination favored by the flux decrement distribution of Ly$\alpha$
forest lines.  Assuming a faint-end luminosity  function slope of 1.58
and  the SDSS estimates of the  bright-end slope and normalization, we
find that the luminosity function must break at $M_B^* = -24.2, -22.3,
-20.8$ at $z=3, 4, 5$ if we adopt the lower ionization rate and assume
no  stellar contribution to the background.   The breaks must occur at
$M_B^*= -20.6, -18.7, -18.7$ for the proximity effect estimate.  Since
stars  may also contribute  to the background,  these values are lower
limits  on the  break luminosity, and   they brighten  by  as much  as
$\sim2$ mag if the  escape fraction  of  ionizing photons from  high-z
galaxies is consistent  with recent estimates: $f_{\rm esc}=0.16$.  By
comparing our expectations to faint AGN searches in the HDF and high-z
galaxy  fields, we   find  that   typically-quoted proximity    effect
estimates of the background imply an over-abundance of AGN compared to
the faint counts (even with $f_{\rm esc}=1$).  Even adopting the lower
bound  on proximity effect measurements,   the stellar escape fraction
must be  high:   $f_{\rm  esc}  \gsim 0.2$.   Conversely,   the  lower
flux-decrement-derived   background  requires   a  smaller  number  of
ionizing  sources, and   faint  AGN counts  are   consistent with this
estimate only if there is a limited stellar contribution, $f_{\rm esc}
\lsim  0.05$.   Our derived  luminosity   functions together with  the
locally-estimated black  hole  density suggest that the  efficiency of
converting mass to light in optically-unobscured AGN is somewhat lower
than expected,  $\epsilon \lsim 0.05$  (all  models).  Comparison with
similar estimates  based on X-ray  counts suggests that more than half
of all  AGN  are obscured in   the UV/optical.  We   also derive lower
limits on typical AGN lifetimes  and obtain $\gsim 10^7$yr for favored
cases. 

\end{abstract}
] 
\section{Introduction}

Among the long-standing goals in extragalactic astronomy is to explain
and characterize the   population  of Active Galactic    Nuclei (AGN).
Their  large luminosities,  compact sizes, and  association with radio
jets  have lead to  the assumption that AGN   are powered by accretion
onto supermassive  black  holes (Salpeter 1964; Zel'dovich  \& Novikov
1964;  Lynden-Bell    1969).  Although   this framework     provides a
theoretical starting point,    there are many questions that    remain
largely unresolved.  These  include    explaining the origin   of  the
central  black  holes (e.g.   Eisenstein \&  Loeb  1995; Madau \& Rees
2001; Koushiappas, Bullock, \&  Dekel 2002), understanding the fueling
process,  lifetime, and efficiency of   the central engine (see,  e.g.
Rees 1984;  Koratkar \& Blaes  1999), and, ultimately, determining how
quasar activity fits within   our  cosmological theory for   structure
formation. 

For many years, the role of AGN in structure formation was believed to
be  that of a tracer  population,  important in  their own right,  but
cosmologically interesting mainly  for  their contribution to  the  UV
ionizing background   (and in their ability  to  track the collapse of
structure).  Recent indications  have changed  this view dramatically.
It now seems likely  that AGN play an important  role in the formation
of galaxies.  In a reversal of sorts, this paper  focuses on using the
observed ionizing  emissivity at high-redshift  in  order to constrain
the evolution of  the AGN luminosity  function.  Derived in this  way,
our luminosity functions relate  directly to the  long-standing desire
to   pinpoint the dominant  ionizing   sources  in the  Universe,  and
additionally help constrain models that  attempt to explain AGN within
a cosmological context. 

The AGN  luminosity  function  has  long  served as  a  benchmark  for
understanding  the  formation and   evolution of quasi-stellar objects
(QSOs)\footnote{We use   the terms  AGN  and QSO  interchangeably.  In
common parlance,  a QSO is  a high luminosity AGN ($M_B\lesssim-23$).}
(Efstathiou  \& Rees  1988;  Carlberg  1990;  Haehnelt  \&  Rees 1993;
Cavaliere, Perri \& Vittorini 1997; Haiman \&  Loeb 1998; Richstone et
al.~1998, Haehnelt, Natarajan,  \&  Rees 1998; Cattaneo, Haehnelt,  \&
Rees 1999;  Haiman,  Madau, \& Loeb  1999, Kauffman  \& Haehnelt 2000;
Haiman \&  Hui 2001; Haehnelt   \& Kauffman 2001;  Steed, Weinberg, \&
Miralda-Escud{\'{e}} 2002).  Recent indications  that AGN activity  is
linked in a fundamental way with the  formation of galaxies make these
studies  all the  more   relevant  (Heckman  et  al.~1984; Sanders  et
al.~1988; Kormendy  \& Richstone 1995;  Sanders \& Mirabel 1996; Boyle
\&  Terlevich 1998;  Dickinson  et  al.~1998;  Magorrian et  al.~1998;
Richstone  et al.~1998; Laor  1998; Wandel 1999;   van der Marel 1999;
Franceschini  et al.~1999;  Mathur 2000;  Canalizo  \& Stockton  2001;
Levenson, Weaver, \& Heckman 2001; Ferrarese 2002).  Specifically, the
relation between black   hole   mass and  bulge   velocity  dispersion
(Gebhardt et al.~2000a,  ~2000b; Ferrarese \&  Merritt 2000; Ferrarese
et al.~2001)   is so tight   that it  seems impossible  to  understand
without some significant cross-talk  (in the form of feedback) between
the AGN phase and the formation of the galaxy and its stellar bulge.

Modelers attempting to understand these  relations have been forced to
test and refine their assumptions by comparing to the low-redshift AGN
luminosity function, or   to bright  quasar  counts at  high-redshift,
because the  faint  population of  AGN is relatively  unconstrained at
high-$z$.  This lack  of knowledge about low-luminosity objects allows
considerable, unwanted freedom for model builders. For example, Haiman
\& Loeb (1998) have explored  the idea that faint  AGN are linked in a
simple way with low-mass cold  dark matter (CDM) halos.  This predicts
a  large number of   low-luminosity systems at  high-$z$  because less
massive halos are   relatively   abundant at  early   times.   Another
possibility is   that AGN activity  in  small halos is   suppressed by
feedback processes.\footnote{ For   example, the binding energy  of  a
galaxy of mass $f_b M$ in a halo of  mass $M$ and circular velocity $V
\propto M^{1/3}$ will  scale  as $E_{\rm   gal} \simeq 0.5  f_b  M V^2
\propto M^{5/3}$.  Compare this to the energy released by a black hole
of mass $M_{\bullet} \propto M$ shining for a time  $t_{\rm agn}$ at a
fixed fraction $\lambda$  of its Eddington luminosity $L_{Edd} \propto
M_{\bullet}$:  $E_{\rm agn} =  \lambda  L_{E} t_{\rm agn}  \propto M$.
This gives $E_{\rm agn}/E_{\rm gal} \propto M^{-2/3}$, suggesting that
AGN feedback  should  be  more important for   low-mass  halos.  If we
insert  appropriate  numbers, we find  that  the energy  released by a
bright AGN  over its  lifetime  should be  comparable  to  the binding
energy of a galaxy-sized halo: $E_{\rm agn} \simeq 1.5 \times 10^{16}$
$M_{\odot}$      km$^2$    s$^{-2}$ (M$_{\bullet}/10^{8}$ $M_{\odot}$)
($t_{\rm agn}/10^{7}$ yr) ($\lambda/0.1$), while $E_{\rm gal} \simeq 2
\times 10^{15}$ $M_{\odot}$  km$^2$ s$^{-2}$ (M/10$^{12}$ $M_{\odot}$)
(V/200  kms$^{-1}$)$^2$ (f/0.1).  Although  just how this energy might
manifest itself as a suppression  mechanism is unclear, the energetics
suggest that a significant amount of feedback is plausible.}  A recent
example of  this idea is explored  in Kauffman \& Haehnelt (2000), who
utilize a qualitatively plausible feedback  scheme to model black hole
properties  and to match the observed  evolution in AGN number density
from $z \sim 0$ to $z \sim 2$.  Because their model relies on feedback
that scales with host population  and redshift, a high-$z$  constraint
on the number of dim objects would serve as a useful test. 

The  AGN   luminosity   function  (LF)   is   typically    written  as
$\phi(L,z)dL$, and is defined   as  the number  of objects  per   unit
comoving volume  at  redshift  $z$,  with luminosity  between  $L$ and
$L+dL$.  In the optical, the majority of studies use B magnitudes.  So
unless otherwise stated, $L$ will denote  B band luminosity throughout
this paper.  For low-redshift  AGN,  $\phi$ is  well represented  by a
broken power law 

\beq       \phi(L,z)       =          \frac        {\phi_*/L_*       }
{(L/L_*)^{\gamma_f}+(L/L_*)^{\gamma_b}}, 
\label{eq:diff_phi}
\eeq 

\noindent 
which has a break at luminosity $L_*$, a characteristic number density
$\phi_*$,  and  asymptotic faint and bright  slopes  of $\gamma_f$ and
$\gamma_b$  respectively.    As  will  be discussed  in    \S2, out to
$z\sim2.5$ the AGN LF seems to evolve only in luminosity, in the sense
that $L_*$  gets  brighter   with  increasing $z$,  while    the other
parameters  stay fixed ($\gamma_f \simeq  1.6$, $\gamma_b \simeq 3.4$,
and $\phi_* \simeq 10^3$Gpc$^{-3}$).  This kind  of evolution is known
as pure luminosity evolution (PLE), and  the current best-estimate for
$L_*(z)$  under this assumption has   it rising dramatically from  its
local value  of $\sim   10^{11}L_{\odot}$  at $z=0$  to nearly   $\sim
10^{13}L_{\odot}$ at  $z = 2.5$ (Boyle,  Shanks, \& Peterson 1988; Koo
\& Kron 1988;   Hewett et al.~1993;  Pei  1995a;  Boyle  et al.~2000).
Beyond this redshift, only the brightest quasars have been observed in
significant numbers, and there is as of yet no evidence for a break in
the luminosity function.   In terms of the  double power-law (1),  the
high-$z$  observations   only measure the  bright-end  slope $\gamma_b
\simeq 2.6-2.9$ (Schmidt, Schneider  \& Gunn 1995, hereafter  SSG; Fan
et al.~2001a) and fix an overall integrated normalization that roughly
imposes  a constraint on the  quantity $\phi_*  L_*^{\gamma_b-1}$ as a
function  of z (see  \S 2).  Most  interestingly, the space density of
the observable bright quasars falls steadily from $z \sim 3$ out to $z
\sim 6$ (Warren, Hewett, \& Osmer 1994 [WHO94]; Kennefick, Djorgovski,
\& de Carvalho 1995 [KDC95]; SSG; Fan  et al.~2001a), but this decline
cannot be  faithfully represented in terms  of PLE,  as the data shows
$\gamma_b$ is flatter at early times.  This leaves us at high redshift
without   a   natural extension  of   the   bright-end  LF to  fainter
magnitudes.  Presumably, future QSO surveys will detect a break in the
LF  at $z \gsim 3$  and measure a faint-end slope.   Until then, it is
useful to examine other constraints. 

A  popular technique  for    constraining a population of   unresolved
sources is to  set an upper limit  based on their contribution  to the
diffuse background   light.    For instance,   AGN are   strong  X-ray
emitters, so measurements of the cosmic X-ray background might be used
to provide upper bounds on the density of AGN.   The problem with this
is that   the  X-ray  background is   measured  locally,  and   we are
interested in constraining a high-$z$ population that contributes only
a small fraction to the $z=0$ signal  (see e.g.  Hasinger 2002).  What
is  preferable is a  measurement of some background at  high $z$ by an
indirect method.   It turns out  that the  UV ionizing  background  is
ideal for  this purpose.  As discussed  in \S 3, the hydrogen ionizing
emissivity  can be measured at   high-redshift by studying Lyman alpha
absorbers    along  the  line of sight    to  distant  quasars.   This
measurement is especially  useful because the  derived background at a
specific  redshift   $z$  is  roughly local:    there   is very little
contribution  from higher redshift sources  because the mean free path
to photo-electric    absorption  is  short compared   to  cosmological
distances (see Madau, Haardt, \& Rees 1999 [MHR] and our Appendix).

\begin{figure}[t]
\centerline{\epsfxsize=8.5cm \epsfbox{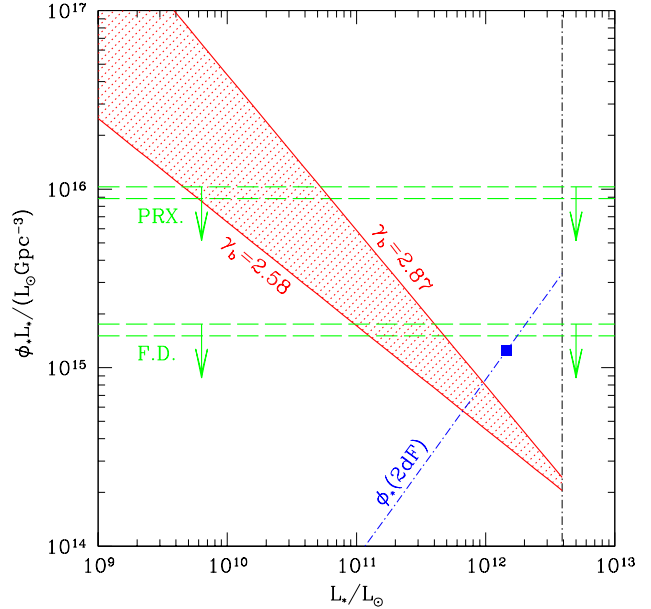}} 
\caption{ Schematic representation of the limits on $\phi_*$ and $L_*$
for $z=4$.  The  integrated   LF ($\Phi(<M_B)$)  basically  constrains
$\phi_*L_*^{\gamma_b-1}$.   We assume   that  all  reasonable  choices
(shaded region) for the  bright-end slope are  bracketed by the values
from  SSG  (2.87) and  SDSS  (2.58).   Since the  observations at high
redshift   show no sign   of a break   for $M_B\lesssim-26$, we do not
consider  this area of   parameter space (vertical  line).  The filled
square is    the point in parameter space    consistent with  both the
integrated LF and the  parameters from 2dF (i.e.   $\gamma_b=3.41$ and
$\phi_*=1070\rm{Gpc}^{-3}$).   We also  bring   in  upper limits  from
ionizing background measurements: proximity effect (``PRX.'') and flux
decrement (``F.D.'').  }\label{fig:phiL_Lz4} 
\end{figure}

%\pspicture(0,0)(8,12.5)
%\rput[tl]{0}(0,12.9){\epsfysize= 8.cm 
%\epsffile{phiL_Lz4.ps} }
%\rput[tl]{0}(0.0,5.){ 
%\begin{minipage}{8.cm} 
% \small\parindent=4.5mm {\sc Fig.}~1. 
% Schematic representation of the limits on
% $\phi_*$ and $L_*$ for $z=4$.  The integrated LF ($\Phi(<M_B)$)
% basically constrains $\phi_*L_*^{\gamma_b-1}$.  We assume that all
% reasonable choices (shaded region) for the bright-end slope are
% bracketed by the values from SSG (2.87) and SDSS (2.58).  Since the
% observations at high redshift show no sign of a break for
% $M_B\lesssim-26$, we do not consider this area of parameter space
% (vertical line).  The filled square is the point in parameter space
% consistent with both the integrated LF and the parameters from 2dF
% (i.e.  $\gamma_b=3.41$ and $\phi_*=1070\rm{Gpc}^{-3}$).  We also
% bring in upper limits from ionizing background measurements:
% proximity effect (``PRX.'') and flux decrement (``F.D.''). 
%\label{fig:phiL_Lz4}
%\end{minipage} 
%} 
%\endpspicture 

In what follows we use this idea to construct AGN luminosity functions
that reproduce the ionizing  background measurements, allowing in some
cases for significant contributions from other  sources (e.g.  stars).
We can    demonstrate  our   general   program using     some   simple
approximations.   Let  us assume (as  we   do throughout) that  we can
approximate each AGN  as emitting light  with  a self-similar spectrum
that varies only in normalization from object to object (\S2.2).  This
implies a fixed ratio between an AGN's specific luminosity at the $912
\mbox{\AA}$  Lyman edge to  its  luminosity in the B-band: $\eta_{\nu}
\equiv L_{\nu}/L  \simeq 10^{18}$ erg/s/Hz/L$_{\odot}$.   The comoving
emissivity at the Lyman  edge is then: $\varepsilon_{\nu} = \eta_{\nu}
\ell$, where $\ell$ is the luminosity density of  AGN (in the B-band).
For a   reasonable LF, $\ell$  will  be dominated  by objects near the
break, so we expect $\ell \sim \phi_* L_*$, or $\varepsilon_{\nu} \sim
\eta_{\nu}\phi_*  L_*$ .  A slightly more  accurate estimate of $\ell$
comes from integrating  the LF (1) from   $L=0$ to $\infty$, and  with
this   we  obtain $\varepsilon_{\nu}   \simeq  \eta_{\nu}   \phi_* L_*
[(2-\gamma_f)^{-1} + (\gamma_b-2)^{-1}]$.   Thus, a measurement of the
ionizing emissivity at  some redshift mostly  constrains the parameter
combination $\phi_* L_*$, with a weak dependence on  the LF slopes (as
long as neither slope  approaches  $2$).  Direct observations, on  the
other hand, measure the   bright-end   slope $\gamma_b$ and  fix   the
normalization of the bright-end tail.  Together, then, the limits from
$\varepsilon_\nu$ and direct observations can effectively restrict the
acceptable values of $\phi_*$ and $L_*$. 

A schematic illustration of  this is shown in Figure  1, where we plot
the available parameter space   at $z=4$.  Direct observations  of the
density of bright  QSOs leads to a constraint  of $\phi_* L_*  \propto
L_*^{2- \gamma_b}$ (see \S  \ref{sect:AGNLF}), which we draw (diagonal
lines) for  two values of $\gamma_b$ from  separate surveys (2.87 from
SSG; 2.58 from  Fan et al.~2001a).  We presume  that the best estimate
of the bright-end  slope   lies between these two   measurements,  and
therefore, that  the  values  of  $\phi_*$   and  $L_*$ are   situated
somewhere in the shaded region.   The fact that  the break has not yet
been  observed down to   the  limiting magnitude $L_{\rm lim}   \simeq
10^{12.6} L_{\odot}$ of SDSS allows us to set an  upper limit on $L_*$
(vertical line).  We can further narrow down  the parameter space with
two different (conflicting) measurements of the ionizing emissivity at
the same redshift (the higher line comes from the proximity effect and
the  lower  line   comes  from  the flux   decrement distribution,  as
discussed in \S  \ref{sect:UVBM}).   We  draw  these as  upper  limits
(horizontal bands), since contributions to the background from non-AGN
sources will require fewer AGN,  further restricting the allowed range
of $\phi_*L_*$.  The width of the bands reflects the slight dependence
of the emissivity on $\gamma_b$.  As  for the faint-end slope, we have
fixed it to  the  low redshift  value  of 1.58, but we   explore other
values of  $\gamma_f$ in \S  \ref{sect:vary}.  What is evident  in the
figure  is  the  fact   that the emissivity    measurements provide an
extremely useful limit on the  parameter space left available from the
direct observations of bright QSOs. 

In  the next   section  we review what   is  known about  the  optical
luminosity   function  from  direct   observations, and   discuss some
possible caveats  associated with dust  and gravitational lensing.  We
also discuss our  assumed template AGN spectrum.  In \S\ref{sect:UVBM}
we  review  the different  ionizing   background measurements, and  in
\S\ref{sect:ionsource},  we  describe   our  calculation of  the   AGN
contribution to the background as well  as the contribution from stars
and  IGM reemission.  We present our  results in \S \ref{sect:results}
for   three separate  models of   AGN emissivity, and  we compare  the
derived    LFs   to past   and   future     faint  surveys.   In    \S
\ref{sect:Implics} we  discuss  our results in  the  context of  other
constraints on the relative AGN and  stellar emissivities at high-$z$.
We also examine what  our results imply for  the AGN  efficiencies and
lifetimes.  Our conclusions are summarized in \S 7 

Unless otherwise stated, we assume that the cosmology  is one with the
universe   made  flat  by a   cosmological    constant $\Omega_m =  1-
\Omega_{\Lambda} = 0.4$,  and that  the  Hubble parameter  at $z=0$ is
$H_0 = 65$ km s$^{-1}$ Mpc$^{-1}$.

\section{AGN Properties: Composite Spectra and Observed Luminosity Function}

In this section we will discuss  the empirical AGN luminosity function
(LF) and  go  on  to present  our  assumed  template  spectral  energy
distribution, which allows us to calculate the ionizing emissivity for
any given $\phi(L,z)$.  As mentioned previously,  $L$ will always have
units  B-band solar luminosities.  We assume  that the  spectrum of an
AGN varies only in normalization from object to object, and we present
our assumed distribution in  terms  of specific luminosity  $L_\nu$ in
units of erg/s/Hz.

\subsection{Empirical AGN Luminosity Function}
\label{sect:AGNLF}

In \S 1 we  discussed how low-redshift  observations indicate that the
AGN luminosity function is well-described  by a double power law shape
given by Eq. 1.  Although yet to be confirmed by observations, we will
assume that this shape provides a useful characterization of the LF at
{\textit{all redshifts}}.   Under  this premise, the  AGN  LF has four
main parameters  that  must be  described   at each $z$:   $\gamma_f$,
$\gamma_b$, $\phi_*$, and $L_*$. 

\begin{figure}[t]
\centerline{\epsfxsize=8.5cm \epsfbox{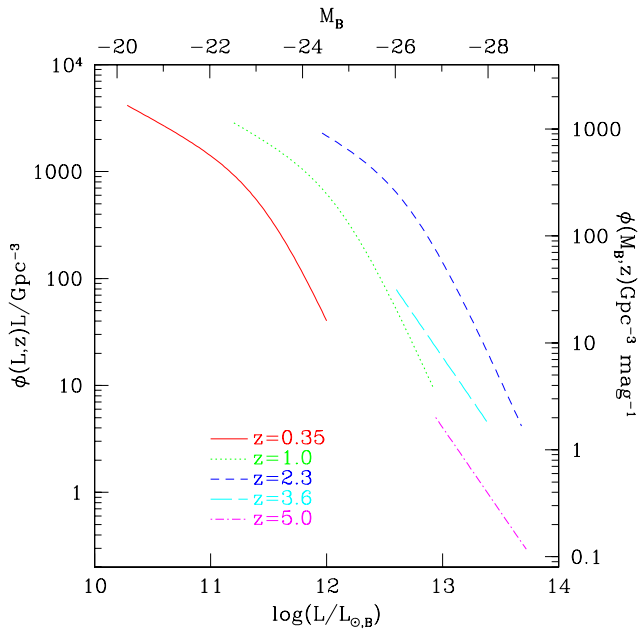}} 
\caption{ Fits  to the observed LF for   several redshifts.  The three
lowest redshift    measurements come  from    Boyle et  al.~(2000) for
$16.50\lesssim M_B\lesssim20.85$.  The  two  highest come from Fan  et
al.~(2001a) for $18\lesssim i^*\lesssim20$.  }\label{fig:phi_obs} 
\end{figure}

Detecting  AGN  at low redshift  is   done primarily by  selecting for
strong UV emitters (e.g. Hartwick \& Schade 1990; Hewett et al.~1995).
The largest survey  to date is that  of  the 2dF (Boyle  et al.~2000),
which  tracks   the  evolution of   QSOs   over  the  redshift   range
$0.35<z<2.3$.  Their data  is  consistent with the assumption  of pure
luminosity evolution (PLE).   Under PLE,  all  of the LF evolution  is
contained  in  $L_*$, while  the  shape ($\gamma_f$,   $\gamma_b$) and
normalization     ($\phi_*$)   remain  constant.    Their  fit   for a
$\Lambda$CDM    cosmology  has    $\gamma_f=1.58$,    $\gamma_b=3.41$,
$\phi_*=1070   \rm{Gpc}^{-3}$,    and   $\log_{10}(L_*)   =   11.24  +
1.36z-0.27z^2$.  The rapid  rise in $L_*(z)$ characterizes an increase
in the number density of  AGN as we  look to higher  $z$.  We plot the
2dF fit  in   Figure \ref{fig:phi_obs}   for several  redshifts.   For
reference, we have only plotted the  fits over the range of (absolute)
luminosities probed    by     the 2dF.    The   rapid    evolution and
characteristic   LF shape found   by the   2dF  for  $z \lsim 2.5$  is
consistent  with previous  work in   overlapping redshift ranges  (see
e.g. Pei 1995a). 

Beyond $z\sim2.5$, however, our knowledge about  the AGN population is
much less complete.  The information we do have is based on the bright
end of the LF, which is observed to fall off  gradually out to $z\gsim
5$ (WHO94, SSG,   Fan et al.~2001a,b).    Of course, if  PLE holds  to
high-$z$ then the evolution of the  brightest quasars is sufficient to
define   the   entire  LF evolution.   For    example,  under the  PLE
assumption,  Pei  (1995a) and later  MHR  used the observed bright-end
evolution  (along with low-$z$  data compilations)  to extrapolate the
evolution of the AGN LF out to high-$z$. Models of this kind have been
popular for estimating the number of high-redshift AGN and calculating
expected AGN contribution to the ionizing background.  Yet even before
the  most  recent SDSS   data, there  were  indications  that  the PLE
assumption    might  break down  at  early   times.  Specifically, the
bright-end slope obtained  by SSG  (and  by KDC95 at slightly   larger
magnitudes) is  flatter  ($\gamma_b\simeq2.9$) than that  measured for
local AGN.   The recent SDSS  data  (Fan et al.~2001a) reveal  an even
flatter bright-end slope ($\gamma_b\simeq2.6$) for  $3.5 \lsim z \lsim
5.0$.  And although  not   as statistically significant, the   data at
slightly fainter magnitudes from the Isaac Newton Telescope Wide Angle
Survey  (Sharp et al.~2001) are  consistent with the results from SDSS
only if  $\gamma_b\lsim2.9$.  All together, these observations provide
strong evidence that the LF evolves in shape as well as in luminosity,
and therefore PLE  does not extend   to high redshift.   In this case,
constraints on the faint-end of the LF become all the more valuable. 

\begin{figure}[t]
\centerline{\epsfxsize=8.5cm \epsfbox{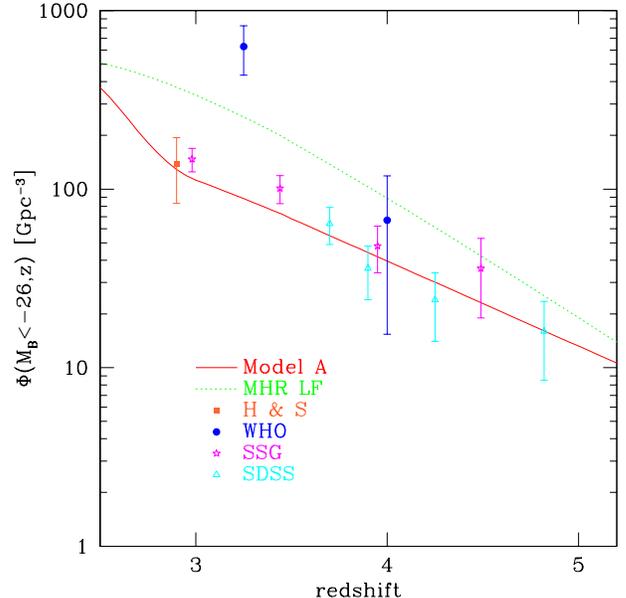}} 
\caption{ The number  density of bright  quasars for $z\gtrsim3$.  The
solid line is one of our models constrained to fit  the SDSS data, and
the dot-dash  line is the PLE LF  from MHR.  The  data points are from
Hartwick \&  Schade  (1990), Warren,  Hewett,  Osmer  (1994), Schmidt,
Schneider, Gunn (1995) and SDSS (Fan et al.~2001a).   Since all the of
the compiled data sets assume $\Omega_M=1$, we  have scaled our fit to
match  this cosmology.   Note  also that  for  Hartwick \&  Schade and
Warren, Hewett, Osmer the error  bars are summed over the differential
LF    uncertainties    and,  therefore,     are  likely overestimated.
}\label{fig:MHRvdata} 
\end{figure}

We will normalize all  of  our high-$z$ LF models   to match the  SDSS
results (Fan  et  al.~2001a,b) over the range  in  $L$  that  they are
measured.   The best-fit LF of   Fan et al.~is  shown  in Figure 2 for
$z=3.6$ and $5.0$ over the relevant range of absolute magnitudes.  The
best-fit bright-end slope  is $\gamma_b =  2.58$ (the quoted error  is
$\pm 0.23$) and the number  density evolution over the redshift  range
$3.5 \lsim z \lsim 6.0$ follows the integrated constraint 

\beq \Phi(M_B<-26,z) = 10^{1.99 -0.47(z-3)} \ \rm{Gpc}^{-3}, 
\label{eq:Phi26_SDSS}
\eeq 

\noindent where

\beq  \Phi(<M_B,z)=\int^\infty_{L(M_B)}\phi(L,z)dL, \label{eq:Phi_def}
\eeq 

\noindent 
and $\log L(M_B)=-0.4(M_B-M_{\odot,B})$.   Note that for  $\gamma_b>1$
and  $L(M_B)\gg L_*$  the integral (\ref{eq:Phi_def})  approximates to
$\simeq \phi_*(L_*/L_m)^{\gamma_b-1}/(\gamma_b-1)$.  The parameterized
normalization from the SDSS  work is shown  in Fig.~\ref{fig:MHRvdata}
along with data from several other high  redshift surveys.  Except for
the  seemingly anomalous point  from  WHO94  at  $z  = 3.2$,  all  the
integrated  LF  measurements   appear  to  agree.   The  figure   also
illustrates how  the PLE model of MHR  disagrees with the recent data,
and highlights  the  need to  reconsider the  AGN  contribution to the
ionizing background at high-$z$.  In what follows we will use the SDSS
result to   fix  the  bright-end  slope   and   normalization  of  our
constrained AGN LFs. 

Before going on, we should   address possible selection effects  which
could affect  the empirical LF at high  redshift.  One  possibility is
that dust, both  intrinsic to the host galaxy  and in  the foreground,
may  be affecting the observed  evolution as well  as shape of the AGN
LF.  For   example, Fall \& Pei (1993)   showed that damped Ly$\alpha$
systems (DLAs) could be blocking 10-70\% of the bright optical QSOs at
$z\sim3$.  But the CORALS survey (Ellison et  al.~2001) found that the
distribution   of  DLAs was  basically the   same  in  both  radio and
optically selected QSO samples.    Since dust should not  affect radio
wavelengths,  this implies that intervening  dust is not a significant
bias.  Moreover, Fan et al.~(2001a)  find the distribution of spectral
indices in the SDSS sample  is similar to that  at low redshift, which
would be unlikely if there was a substantial amount  of dust along the
line of sight.  So too, any increase  in reddening with redshift would
presumably be  at odds with the  close agreement in LF  evolution from
surveys with very different selection techniques: the broad-band color
selection in SDSS; the Ly$\alpha$-emission selection  of SSG; and even
radio selection (e.g.  Hook, Shaver \& McMahon  1998), which should be
representative  of full sample,  assuming that the ratio of radio-loud
to radio-quiet  QSOs is   redshift  independent (Stern  et  al.~2000).
However,  none of   this would  seem to  preclude  the  existence of a
\emph{distinct}  population    of   highly  extinguished   QSOs,  only
observable in  the X-ray (and perhaps  the IR).  We  come back to this
possibility when we    address  the relic    black  hole  density   in
\S\ref{AGN_BH_densities}. 

Another effect that could distort the LF is gravitational lensing (see
Blandford \&  Narayan  1992 and references   therein).  Because of the
shape of the LF, the magnification bias  is stronger for more luminous
objects, i.e.  it is   more likely that  the  bright-end of the LF  is
contaminated  by  artificially   brightened   AGN.  According  to  Pei
(1995b), the number of objects observed with $L=10L_*$ that are lensed
could be as much as  44\% at $z=3$ and  68\% at $z=5$.  However, these
large lens  fractions come from models that  assume a now unreasonable
density   of compact  objects  (on   the order of   10\%  the critical
density).  More recent estimates of  the fraction of lensed objects in
magnitude limited surveys  are less than  $\sim20\%$  (Barkana \& Loeb
2000;   Wyithe \& Loeb   2002).  Because  these  are relatively  small
corrections, we  will ignore the effect  of magnification  bias in our
bright-end normalizations.  However, it should be pointed out that the
lens fraction gets larger the  farther an observed luminosity is above
the break.  That  is, the  probability  that the QSOs observed  in the
SDSS ($L\gtrsim10^{12.5}\rm{L_{\odot,B}}$)  are lensed  increases  the
smaller $L_*$  turns out to be.   In our emissivity constrained models
(see Figure  \ref{fig:phiL_Lz4}),  we explore break luminosities  that
are  several   orders of  magnitude below   those  investigated by Pei
(1995b).  However,   these values  of $L_*$ are   probably unrealistic
because  they seem   to be in   conflict with  some of   the faint AGN
searches (see \S \ref{sect:results}).

\subsection{Lyman Limit emissivity and Composite Spectra}

\begin{figure}[t]
\centerline{\epsfxsize=8.5cm \epsfbox{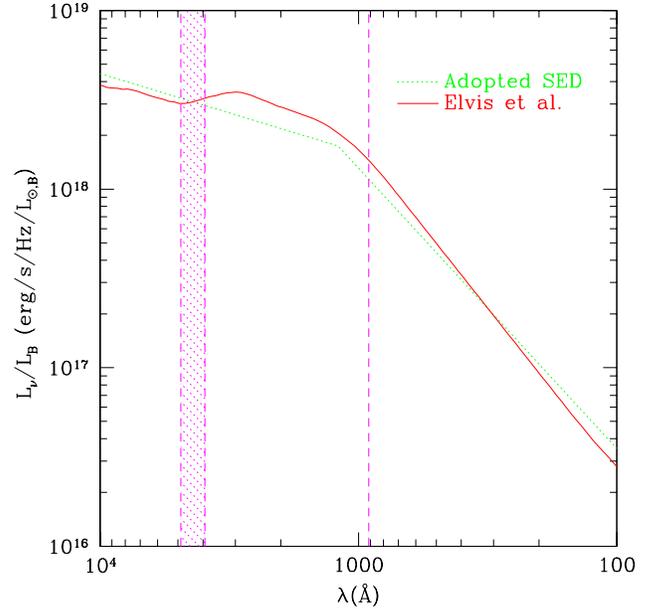}} 
\caption{  Our  assumed template  AGN   spectrum shown along  with the
composite spectrum  of Elvis  et al.~(1994).   For reference,  we have
drawn vertical lines at the Lyman limit and a  shaded band to indicate
the wavelength range of the B-band.  }\label{fig:SED} 
\end{figure}

Given an optical AGN LF $\phi(L,z)$, we are concerned with calculating
the implied emissivity  of ionizing photons.   If $\varepsilon(\nu,z)$
is the comoving emissivity of photons  at frequency $\nu$ and redshift
$z$, then we can write 

\beq   \varepsilon(\nu,z)  =   \int^{\infty}_{L_{\rm  min}} dL   \cdot
\phi(L,z) \cdot L_\nu(L,\nu). 
\label{eq:emissivity} 
\eeq 

\noindent 
Here  (and  throughout this work)  we adopt  a  minimum AGN luminosity
$L_{\rm min} =10^8 \rm{L_{\odot,B}}$ (or $M_B=-14.5$).  This choice is
comparable to the nuclear luminosity of the faintest Seyferts (Londish
et   al.~2000).  Fortunately, the   choice  of $L_{\rm min}$ does  not
strongly  affect  our  results     as   long as  $L_{\rm     min}  \ll
L_*$.\footnote{If we let  $L_{\rm min}=10^9 \rm{L_{\odot,B}}$  we find
it affects our results   (\S 5) by   less than 30\%.  And  making  the
minimum smaller than our fiducial  value has almost  no effect, as the
integrals converge.}   Expression \ref{eq:emissivity}  provides a link
between  the  emissivity and $\phi(L,z)$,  once  we specify  the input
spectrum, $L_{\nu}(L,\nu)$.  We   will assume  that, on  average,  the
specific  luminosity of an  AGN  (as  a function  of frequency  $\nu$)
varies only in normalization  such that $L_{\nu}  \propto L$, with  no
$z-$dependence (see Kuhn et al.~2001).

From the UV to optical, an AGN spectrum, $L_{\nu}$, is reasonably well
approximated  by a double  power law  with a break  near Ly$\alpha$ at
$1216 \mbox{\AA}$.  Since the Galaxy is  opaque to UV photons, we must
rely on  high redshift quasars  to determine a typical quasar spectrum
short-ward of the  Lyman limit.  For  $\lambda < 1200  \mbox{\AA}$, we
will use results from  the most recent UV  survey from HST  (Telfer et
al.~2002).  Telfer and collaborators find that the specific luminosity
of (radio quiet)  AGN  scales as  $L_{\nu} \propto\nu^{-\alpha_{UV}}$,
$\alpha_{UV}=1.57\pm0.17$, from  $\lambda   \sim 500$  to $\sim   1200
\mbox{\AA}$.\footnote{ We assume that the minority population of radio
loud objects  will contribute very little  to  background.  It is also
worth  noting that the  Telfer et al.~(2002)  slope is slightly harder
than    previous results from HST  (Zheng    et al.~1997), which found
$\alpha_{UV}=1.83\pm0.15$.  If we rerun our analysis with this steeper
continuum, we find   that  our results   change  by less  than  20\%.}
Long-ward of Ly$\alpha$,   we   use  the  result of  Vanden   Berk  et
al.~(2001), who relied on a composite of  quasar spectra from the SDSS
to determine $L_\nu\propto\nu^{-0.44}$ from $1200 \mbox{\AA}$ to $\sim
5000 \mbox{\AA}$.

Figure  \ref{fig:SED}  shows our  assumed   spectrum  compared to  the
composite spectrum of Elvis et al.~(1994).  The specific luminosity is
normalized  relative to the  B-band   luminosity $L$.  For our  B-band
transformation we follow SSG  and use $M_B = M_{AB}(4400\mbox{\AA})  +
0.12$.   Of course, what we are  mainly interested in  is the ratio of
$L_\nu$ at $912\mbox{\AA}$  ($\nu = \nu_H$)  to the optical luminosity
$L$, which for our spectrum is 

\beq \eta_\nu \equiv \frac{L_\nu(\nu_{\rm{H}})}{L} = 10^{18.05} \ \rm{
  \frac{ergs/s/Hz}{L_{\odot,B}} }. 
\label{eq:Lnu_L} 
\eeq 

\noindent 
From combining    the optical and   UV  power  laws,   we estimate  an
uncertainty in $\eta_\nu$  of $\pm 0.10$ in  the exponent.  This level
of variation will not strongly affect our conclusions.  For reference,
the value implied by Elvis et al.~(1994) is $\eta_\nu = 10^{18.15}$ in
the same units, and the value obtained  by Shull et al.~(1999) using a
sample of 27 Seyferts is  $\eta_{\nu} = 10^{17.93\pm0.04}$.  Note that
Shull et  al.~also find a slight  luminosity dependence on this ratio,
but  we will ignore that possibility  here.  With  our assumed $L_\nu$
spectrum,  Expression   \ref{eq:emissivity} fully     describes    the
relationship between a given optical LF and the ionizing emissivity as
a function of redshift.

\section{Ionizing Background Measurements}
\label{sect:UVBM}

From the Gunn-Peterson effect  (Gunn \&  Peterson  1965), it is  clear
that the IGM is  highly ionized.  If it can  be assumed that this high
ionization state is due to photo-ionization (as opposed to collisional
excitation), then measurements of the IGM, particularly the Ly$\alpha$
forest, can  reveal  information  on    the nature of    the  ionizing
background.   We can  characterize this  background  with the hydrogen
ionization rate: 

\beq       \Gamma(z)   =   4\pi     \int^\infty_{\nu_{\rm{H}}}    d\nu
 \frac{J(\nu,z)}{\nu}\sigma_{\rm{H}}(\nu).  \eeq 

\noindent 
Here  the cross section    to  hydrogen photo-electric  absorption  is
$\sigma_{\rm{H}}=6.35\times10^{-18}\rm{cm}^2\cdot(\nu/\nu_{\rm{H}})^{-3}$
and   $J(\nu,z)$  is     the   background intensity    (in  units   of
$\rm{ergs/s/cm^2/Hz/sr}$).  As will be discussed  in the next section,
the  ionizing intensity  relates    simply to  the AGN   (and stellar)
emissivity, $J(z) \propto \varepsilon(z)$,  so we may  approximate the
spectrum  of background  intensity  as   $J\propto\nu^{-\alpha_{UV}}$.
With this approximation we obtain: 

\beq \Gamma_{-12}(z) = \frac{12.0}{3+\alpha_{UV}} \ J_{-21}(z) 
\label{eq:G_12_J_21}
\eeq 

\noindent 
where   $J_{-21}(z)\equiv       J(\nu_{\rm{H}},z)$  in     units    of
$10^{-21}\rm{ergs/s/cm^2/Hz/sr}$   and    $\Gamma_{-12}         \equiv
\Gamma/10^{-12}\rm{s^{-1}}$.  If there  were  only a  single  dominant
source   to    the  background (e.g.    QSOs),    then  we  could take
$\alpha_{UV}$ to be equal to the typical spectral index of that source
population.  We will however be  exploring cases where there are  more
than one  dominant type of  emitters, and so   we will not immediately
assume a value of  $\alpha_{UV}$.\footnote{Moreover, the processing of
the background through absorption and reemission in the IGM will alter
its spectral shape (Miralda-Escud{\'{e}} \& Ostriker 1990).}  For this
reason,  in the next  section,  we   will  quote measurements of   the
ionizing background using $\Gamma_{-12}$ (instead of $J_{-21}$). 

Under the assumption  of photo-ionization equilibrium, $\Gamma$ should
scale     as   the  ratio  of       ionized    to  neutral    hydrogen
($n_{\rm{HII}}/n_{\rm{HI}}$).  Given this ratio one  may infer a value
for $\Gamma$.  At  high-$z$, two  techniques  exist for measuring  the
amount   of neutral hydrogen, and  they  both rely on using Ly$\alpha$
lines in the  spectra  of distant  QSOs.\footnote{  Note, because  the
Ly$\alpha$ forest is very thin for  $z\lesssim1.7$, it is difficult to
make  measurements  of the  ionizing  background  in this  way at  low
redshift (e.g.  Shull et al.~[1999] and references therein).  Although
see Dav\'{e}    \& Tripp (2001).}   One  technique  relies on analytic
modeling  of the expected distribution of  lines in  the vicinity of a
QSO   (the proximity effect)     and the  other  uses  hydro-dynamical
simulations to model the expected distribution of lines along the line
of sight  to  the quasar (what  we  will  call the ``flux  decrement''
analysis).  We use the  following   two subsections to  summarize  the
results of each technique.

\subsection{Proximity effect}
\label{sect:prox}

The  proximity  effect was  first  discussed  by  Bajtlik, Duncan,  \&
Ostriker (1988).  The technique relies on the observed decrease in the
number of Ly$\alpha$ lines  in the vicinity of a  QSO relative to what
one would have expected in the absence of the QSO.  In principle, this
decrease is a result of ionizing radiation from  the QSO itself, which
tends to reduce the neutral fraction  in nearby absorbers.  The number
of observable lines  should decrease correspondingly.   An estimate of
the ionizing rate can be obtained by determining the distance from the
quasar at  which   the number of   lines is  equal  to  the background
expectation.   Over  the  years,  measurements of  $J_{-21}$ (assuming
$\alpha_{UV}\sim1.8$) have  taken values between $\sim0.7$ and $\sim3$
(Williger     et al.~1994;   Bechtold   1994; Fern{\'{a}}ndez-Soto  et
al.~1995; Cristiani  et al.~1995; Giallongo  et al.~1996; Cooke, Espey
\& Carswell 1997;  Scott et al.~2000; Liske \&  Williger 2001).  Since
what  is effectively  measured is  a ratio  of proper distances,  this
analysis should be relatively  independent of cosmological parameters.
Most estimates of the proximity  effect assume that the background  is
constant over the redshifts being sampled, so  this technique does not
appear  to  uniquely  determine  the evolution   of $\Gamma_{-12}(z)$.
Haardt \& Madau (1996) and Fardal et  al.~(1998) have tried to fit the
whole set of proximity effect measurements  with a Gaussian.  Scott et
al.~(2000) claim that their observations are well fit by the following
parameterization from Fardal et al.~(1998): 

\beq
  \Gamma_{-12}(z)=1.2(1+z)^{0.58}\exp\left[\frac{-(z-2.77)^2}{2.38}\right].
  \eeq 

\begin{figure}[t]
\centerline{\epsfxsize=8.5cm \epsfbox{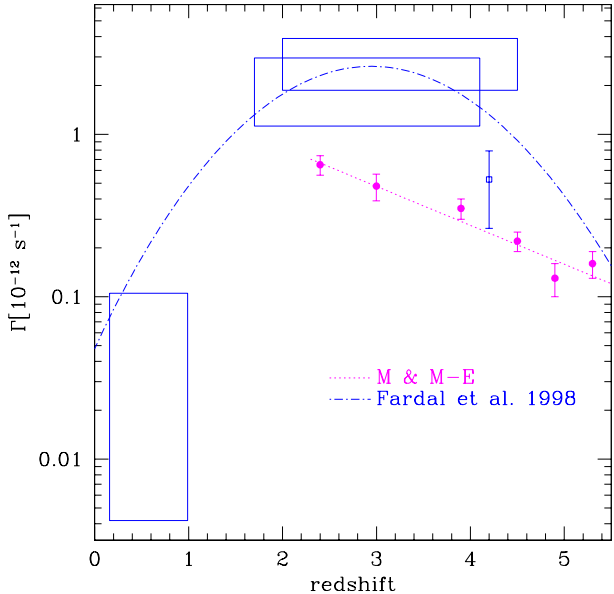}} 
\caption{ The ionizing rate as a function of redshift.  The solid dots
are from hydro-simulations  by McDonald  \& Miralda-Escud{\' e}(2001).
A linear fit  has been drawn through  the points.  The solid boxes and
open square are representative of recent proximity effect measurements
(see  Scott  et  al.~2000),     where we  have used    the  conversion
$\Gamma_{-12}=2.64\cdot J_{-21}$ from Fardal  et al.~(1998).  We  have
also  drawn   the Gaussian   fit to this    data from  Fardal  et  al.
}\label{fig:Gamma_dat} 
\end{figure}

\noindent 
This      fit is represented  by     the   dot-dashed  line in  Figure
\ref{fig:Gamma_dat}. 

There is some concern that the value of $\Gamma_{-12}$ determined from
the proximity effect is  overestimated.    This is because  QSOs   are
likely to be   found  in environments  that are  denser  than  average
(Pascarelle et al.~2001; Ellison et  al.~2002), so that the regions of
excess ionization  end up being smaller  than they would  have been if
the region was of   average density.  Such overdensities would  likely
scale with the  mass/luminosity of the   AGN, which might  explain the
slight anti-correlation between more luminous QSOs and their proximity
effects (see   Cooke, Espey \&  Carswell   1997).  Loeb  \& Eisenstein
(1995) claim this bias could  cause the proximity effect  measurements
to overestimate the background by as much as a  factor of 3.  The flux
decrement technique discussed   in the next  subsection  does  in fact
yield a lower ionizing background than  the proximity effect, as would
be expected if  these concerns are  valid.   However, we will keep  an
open mind about  the  issue and  discuss the constraints  on faint AGN
associated with each technique separately. 

\subsection{Flux Decrement}
\label{sect:fluxD}

Another  way to utilize the Ly$\alpha$  forest to measure the ionizing
background  requires making a theoretical prediction  as to the amount
of (unobserved) ionized hydrogen in  each ``cloud'' along the line  of
sight to a quasar.   This  can be  done  using CDM theory  (and N-body
codes)   coupled  with   hydrodynamical  simulations  that   model gas
evolution (Cen et al.~1994; Zhang et  al.~1995; Hernquist et al.~1996;
Miralda-Escud{\'{e}}  et al.~1996;  Rauch  et al.~1997; Dav{\'{e}}  et
al.~1997;   Wadsley \&  Bond   1996;  Zhang et  al.~1998;  McDonald et
al.~2000).  These simulations  essentially generate  a distribution of
hydrogen  Ly$\alpha$  optical  depths   which  can   be mapped  to   a
distribution of flux  decrements ($F=\exp(-\tau)$).  The distributions
generated in this  way convincingly reproduce observations, adding  to
the long   list of  successes  for the   CDM paradigm.   Although  the
\emph{shape} of the  flux decrement distribution depends upon  details
of   the  cosmology as  well    as models for the  temperature-density
relation, the \emph{normalization}  is proportional only to the baryon
density,  the ionization rate\footnote{It  is usually assumed that the
ionization rate is homogeneous.  Gnedin \&  Hamilton (2002) included a
spatially inhomogeneous radiation  background from  galaxies in  their
simulation and found  that the mean  background came out larger.  This
could mean  that   the flux decrement measurements  underestimate  the
ionization rate by at least  20\%; however, a significant contribution
from AGN would likely lessen this effect.}, and the Hubble parameter: 

\beq \tau \propto \frac{(\Omega_b h^2)^2}{\Gamma(z) \ H(z)}.  \eeq 

\noindent 
Often times this relation is used to put constraints on $\Omega_b h^2$
 (e.g. Hui et  al.~2002).  But McDonald \& Miralda-Escud{\'{e}} (2000)
 [hereafter   M\&M-E]    assume  the    baryon  fraction    from  BBN:
 $\Omega_bh^2=0.02$ (Walker et al.~1991; Burles \& Tytler 1998), so as
 to measure   $\Gamma(z)$   from a  sample   of  QSO  spectra.    In a
 $\Lambda$CDM cosmology, over the redshift range of $\sim2.3-5$, their
 results can be approximated by 

\beq \Gamma_{-12}(z)=10^{-0.24z+0.4}.  \eeq 

\noindent 
We   plot this flux-decrement result   along with the proximity effect
results in   Figure    \ref{fig:Gamma_dat}.    The   flux    decrement
determination is a  factor of $\gsim  4$ below  that of the  proximity
effect.  We  discuss how these differences   affect constraints on the
AGN LF in \S 5.\footnote{ In recent  papers (Cen \& McDonald 2002; Fan
et al.~2002),  the   flux-decrement  analysis  has  been extended   to
$z\sim6$,  and  evidence  for  the  epoch   of reionization   has been
discussed.  If  the epoch of  reionization has been detected, then the
drop in the  ionization rate at  $z \sim 6$ will  be due in part to  a
decrease in the mean free path of continuum photons, and not merely to
a change in the ionizing emissivity.  See \S \ref{sect:ionsource}.} 

\section{The IGM: AGN and stars as ionizing sources}
\label{sect:ionsource}

Now that we have measurements  of  the background ionization rate,  we
need to    relate them to   the LF.   This relation comes    about via
Eq.~\ref{eq:G_12_J_21}  which   connects  $\Gamma$  to  the   ionizing
intensity.  In principle, the intensity at any time is  made up of the
integrated  contribution  of  all  sources over    the history of  the
universe.  However, as discussed   in detail by MHR, the  contribution
from distant sources  is significantly degraded  by  attenuation.  The
intensity can be written as an integral over the (comoving) emissivity
of sources as a function of redshift: 

\beq J(\nu,z)    = \frac{c}{4\pi}  \cdot (1+z)^3  \cdot  \int^\infty_z
 d\bar{z}  \frac{dt}{d\bar{z}}  \cdot   \varepsilon(\bar{\nu},\bar{z})
 \cdot \rm{e}^{-\tau_{\rm eff}(\nu,z,\bar{z})},   \label{eq:intensity}
 \eeq 

\noindent 
where $\bar{\nu}=\nu \ (1+\bar{z})/(1+z)$, and $\tau_{\rm eff}$ is the
effective optical  depth due to  photo-electric absorption in the IGM.
In  the  Appendix we discuss how   we  model the IGM  and $\tau_{eff}$
following the prescription of MHR.  In practice we calculate $J$ using
the full integral expression, but  for purposes of illustration, it is
useful to  make the following approximation.  If  we assume  that only
ionizing sources   within one  absorption   length, $\Delta l$  (where
$\tau_{\rm eff}(\Delta l)  \equiv  1$), contribute to the  background,
then   Eq.~\ref{eq:intensity}   reduces  to  $J(\nu,z)  \simeq (1+z)^3
\varepsilon_\nu(z) \Delta l / 4 \pi$.   At the Lyman edge, our adopted
IGM      model        (Appendix)    gives       $\Delta     l   \simeq
39\rm{Mpc}[(1+z)/4]^{-4.5}$.   Since $\Delta  l   \ll c/H(z)$ this  is
often called the ``local source'' approximation, and it yields: 

\beq      J_{-21}(z)      \simeq      0.17       \  (1+z)^{-1.5}     \
\varepsilon_{24}(\nu_H,z). 
\label{eq:Japprox}
\eeq 

\noindent 
Here  we have introduced  the  symbol $\varepsilon_{24}$  which is the
emissivity in units of 10$^{24}$ ergs/s/Hz/Mpc$^3$. 

Equations \ref{eq:intensity} and \ref{eq:G_12_J_21} allow us to relate
an observed ionization   rate  $\Gamma(z)$ to a    background ionizing
emissivity $\varepsilon(z)$.        However,    the  relation  depends
sensitively on the  spectral slope of  the background: $\Gamma \propto
J/ (3 + \alpha_{UV}) \propto \varepsilon/(3  + \alpha_{UV})$.  Because
the  far-UV slopes  of galaxies and   AGN are significantly different,
deriving a limit on the  background emissivity based on the ionization
rate would  necessitate  assuming   something about  the    background
population.  We would  prefer a more  general parameter that allows us
to  consider  stars  as well as  AGN  as   major  contributors to  the
background.  For this  purpose  we propose the following  ``weighted''
emissivity parameter 
  
\beq                  \hat{\varepsilon}^i                       \equiv
\frac{\varepsilon_{24}^i(\nu_H)}{3+\alpha_{UV}^i}, \eeq 

\noindent 
which facilitates a  fair  comparison between  different  populations,
$i$.  This definition allows  us to rewrite the (approximate) Equation
\ref{eq:Japprox} as 

\beq     \Gamma_{-12}(z)   \simeq  2.0      \ (1+z)^{-1.5}   \  \sum_i
\hat{\varepsilon}^i(z), 
\label{eq:glim} 
\eeq 

\noindent 
where the  sum is over  all populations  of ionizing ionizing sources,
$i=$ AGN,  stars, reemission, etc.   This expression can be rearranged
to place  an upper limit  on the weighted  emissivity  coming from any
individual population.  In  particular, we are concerned with limiting
the emissivity from AGN: 
 
\beq \hat{\varepsilon}^{Q}(z) \   \simeq  \ 0.5 \  \Gamma_{-12}(z)   \
(1+z)^{1.5}    \     -      \          [\hat{\varepsilon}^{*}(z)     +
\hat{\varepsilon}^{R}...]. 
\label{eq:eplim}
\eeq 

\noindent 
The bracketed terms on the right hand side  are included to illustrate
how the  allowed  AGN  emissivity  is  reduced  by  the presence    of
additional     contributing      populations.      For        example,
$\hat{\varepsilon}^{*}$ represents the  weighted emissivity from stars
(see \S \ref{sect:stars})  and  $\hat{\varepsilon}^{R}$ represents the
contribution from  cloud reemission  (see  \S \ref{sect:reem}).    The
contribution from clusters  of galaxies is likely  negligible (Randall
\& Sarazin 2000). 

\begin{figure}[t]
\centerline{\epsfxsize=8.5cm \epsfbox{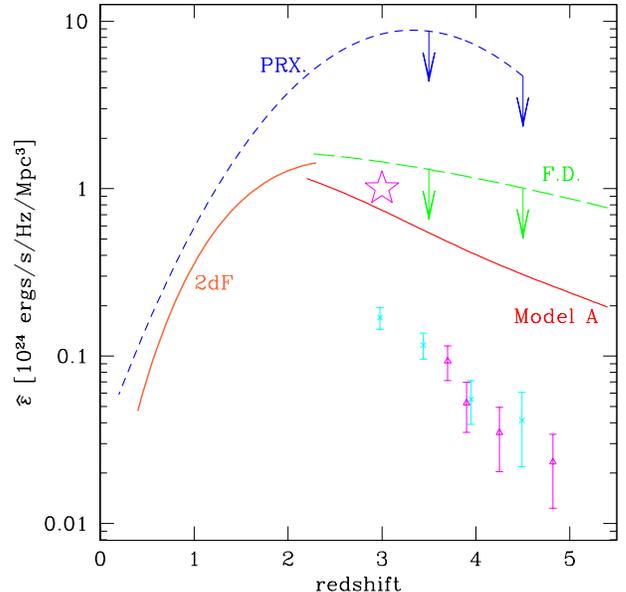}} 
\caption{The                      weighted                 emissivity:
$\varepsilon(\nu_{\rm{H}})/(3+\alpha_{UV})$.  The  bold solid line  is
the weighted emissivity implied  by observed QSOs ($-26<M_B<-23$) from
2dF  (Boyle et al.~2000).  The crosses  are from SSG and the triangles
are from the  SDSS (Fan  et  al.  2001a), both  involving a conversion
from $\Phi(M_B<-26)$.   The star is  the weighted LBG  emissivity from
Steidel et al.~(2001).  Limits have been drawn from the two ionization
rate measurements: proximity effect (PRX.)  and flux decrement (F.D.),
see \S \ref{sect:prox} and \S \ref{sect:fluxD}.  We have also included
our Model  A  (solid  line),  which   is  the allowed AGN   emissivity
accounting  for  a stellar  contribution to the flux-decrement-derived
background (see \S\ref{sect:results}).  }\label{fig:epsilon_hat} 
\end{figure}

Fig.~\ref{fig:epsilon_hat} illustrates the upper limit on weighted AGN
emissivity $\hat{\varepsilon}$ imposed  by the proximity effect (short
dashed  line)  and flux decrement  (long-dashed  line) measurements of
$\Gamma(z)$.   As will be  the case throughout, we  have used the full
integral expression given by  Eq.  11 in order  to place limits on the
weighted emissivity (as  opposed  to the approximate expressions  used
for illustrative purposes in  Eqs.  12, 14, and  15).  However, in the
language of Eq. 15, these two constraints amount to setting all of the
bracketed (non-AGN) emissivities to zero.  The  thin solid line (Model
A), on the other hand illustrates how the flux-decrement-derived limit
on the AGN emissivity changes if  we include a (conservative) estimate
of the stellar  emissivity, $\hat{\varepsilon}^*$.   We describe  this
estimate in the    next   subsection.  For  reference  we    show  the
contribution to $\hat{\varepsilon}$   arising from AGN that  have been
directly observed by the 2dF at low-$z$ (bold solid  line), and by SSG
and the SDSS at high-$z$ (data points). 

We   now go on to  discuss   our stellar emissivity  estimate in  more
detail. 

\subsection{Emission from Galaxies}
\label{sect:stars}

In addition to AGN, star-forming  galaxies are an obvious  contributor
to  the ionizing continuum.   At   energies above  the Lyman edge,   a
star-forming galaxy spectrum is dominated  by hot, short-lived O stars
($t_{\rm   O}\lesssim10^{7}\rm{yr}$).  Theoretically,  the number   of
Lyman continuum photons emitted by an (unobscured) $L_*$ galaxy should
be about $10^{-3}$ smaller than  the number emitted  from an $L_*$ AGN
(Madau \& Shull 1996).  However,  the space density of $L_*$  galaxies
is a factor of $\sim 10^4$ higher than that of $L_*$ AGN at $z\sim 0$,
so  stars  could conceivably dominate  the  ionizing background.  What
mitigates this calculation is the fraction  of Lyman continuum photons
that can actually escape from the galaxy, $f_{\rm esc}$. 

Theoretically, cold gas   in a galaxy could  trap  most  of the  Lyman
continuum  emission   (e.g.   Haehnelt  et   al.~2001   and references
therein).  In agreement with this expectation, most local searches for
Lyman  continuum  emission from galaxies have  come  up empty.  At low
redshift, using  the   Hopkins  Ultraviolet Telescope,   Leitherer  et
al.~(1995) found  that nearby  starburst  galaxies  had  a very  small
escape fraction, $f_{\rm esc} < 3 \%$ (see  also Hurwitz et al.~1997).
Similar upper  limits can be found from  FUSE (Deharveng  et al.~2001)
and from HST  measurements  in the  HDF (Ferguson  2001).  The  escape
fraction    from the  Milky  Way    can be   estimated from  H$\alpha$
measurements   in  the Magellanic  Stream  ($f_{\rm esc}\sim6\%$), but
there   is  some   uncertainty  associated   with  this estimate  (see
Bland-Hawthorn \& Maloney 2001). 

At high  redshift  there are  indications that the  escape fraction is
much  higher   than these   local  observations suggest.    Steidel et
al.~(2001) reported  the  first evidence for galactic  emission beyond
the Lyman edge using a set of galaxies at $z \sim 3$.  Their result is
based   on   a  sample    $\sim  1000$    spectroscopically-confirmed,
star-forming,  Lyman  Break  Galaxies  (LBGs),  for   which the  total
emissivity      at $1500\mbox{\AA}$   is     fairly-well   determined:
$\varepsilon_{24}^*(1500\mbox{\AA})     =  180$    at  $z=3$  for  our
$\Lambda$CDM cosmology   (Steidel  et al.~1999).   Using  29  galaxies
effectively taken from  the bluest quartile  of their LBGs, Steidel et
al.~(2001) derived a composite spectrum, and used it to study the flux
density   at   $1500\mbox{\AA}$  compared   to  $900\mbox{\AA}$.  They
obtained the  surprisingly low value f[1500]/f[900]   $= 4.6 \pm 0.1$,
which is nearly  consistent with theoretical expectations \emph{in the
absence of any internal absorption} ($\sim 3-5$ according to Leitherer
et  al.~1999, Steidel et  al.~2001,  and references therein).  If  one
assumes the intrinsic ratio  is $3$, then  the escape fraction implied
by their  measurement  is $f_{\rm  esc}  =  65\%$.  If  all LBGs  were
emitting ionizing photons at this rate then they would likely dominate
the AGN-contribution to the background, and, in fact, over-produce the
ionizing  rate  implied  by  flux  decrement  measurements.   However,
Giallongo et al.~(2002) recently looked at  2 random LBGs and found no
Lyman  continuum photons.   It  seems  more likely,  or  at least more
conservative, to assume  that  only the bluest   quartile of  the  LBG
population at $z\sim3$ is  emitting ionizing photons with $f_{\rm esc}
= 65\%$, and the rest are much more opaque.  Under the assumption that
only the bluest  fourth  have escaping  Lyman continuum  photons,  the
average escape fraction  for the  entire  population would be  $f_{\rm
esc}= 65\%/4 \simeq 16\%$.  This is what we will assume here. 

In  terms of the escape fraction,  the implied $z=3$ emissivity at the
Lyman  edge  is $\varepsilon_{24}^*(\nu_H)   = (f_{\rm  esc}/3)  \cdot
\varepsilon_{24}^*(1500\mbox{\AA})  \simeq    60 \cdot   f_{\rm esc}$.
However, what we are concerned with is the stellar contribution to the
ionizing rate,  which depends    on   the far-UV   background    slope
$\alpha_{UV}$.   As in Eq.    13, we characterize the slope-dependence
using the weighted emissivity of stars: 

\beq      \hat{\varepsilon}^*   =    \frac{\varepsilon_{24}^*(\nu_H)}{
3+\alpha^*_{UV}} = \frac{60 \cdot f_{\rm esc}}{3+\alpha^*_{UV}}, \eeq 

\noindent
where we  have implicitly assumed $z=3$.   For the slope we will again
rely  on     the   results of     Steidel    et  al.~(2001).   Between
$1100\mbox{\AA}$ and $900\mbox{\AA}$, their spectrum yields a slope of
$\alpha_{UV} \simeq  6.7$, and we will  assume that this power-law can
extrapolated     into the    Lyman    continuum   without  significant
error.\footnote{One might  expect  a particular starburst  spectrum to
show a break at  the Lyman limit, so that  our choice of extending the
Far UV   ($\lambda>912\mbox{\AA}$)   spectral index   into   the Lyman
continuum may be unwarranted.  However,  since, within our picture  at
least, the main contributors have almost no attenuation such an effect
would not be  expected.  For example, it is  possible that  the bluest
LBGs  are being seen through  ``superbubbles'' (Dawson et al.~2002) in
which the neutral   hydrogen has been  partially blown  out or ionized
away.   This would leave   the spectrum  relatively smooth  across the
Lyman    break, whereas  along some other     line of sight, the  same
starburst may  have  almost no Lyman continuum   emission.  In such  a
scenario, $f_{\rm  esc}$ would   absorb these viewing-angle  effects.}
With $\alpha_{UV}=6.7$ and $f_{\rm  esc} = 16\%$, the implied weighted
emissivity  for   stars   at  $z=3$   is $\hat{\varepsilon}^*   \simeq
1.0$. This value is represented by the five-pointed star in Figure 6. 

One  concern is that  the value of $\alpha_{UV}$  we derived using the
Steidel  et al.~(2001)    composite   spectrum is  much   softer  than
theoretical  models   often assume.    A   more common  assumption  is
$\alpha_{UV}\simeq 2$   (e.g.  Miralda-Escud{\'{e}} \&  Ostriker 1990;
Madau \& Shull 1996; Shull et  al.~1999; Haehnelt et al.~2001).  While
our use of  the Steidel slope is well-motivated,  if one were inclined
to believe a  different value, we  could simply  absorb the  change in
$\alpha_{UV}$   by re-interpreting our  assumed   $f_{\rm  esc}$.  For
example,    if     we   adopt   the     theoretically-motivated  value
$\alpha_{UV}=2.0$, then we obtain the same $\hat{\varepsilon}^*$ using
$f_{\rm  esc}\simeq 8\%$.  Since our main  goal  in using this stellar
emissivity model is to  illustrate how it  will affect the implied AGN
LF,  any reader unhappy with  our adopted slope  can simply regard our
assumption to be   $\hat{\varepsilon}^*_{24} \simeq 1$ at   $z=3$, and
interpret $f_{\rm  esc}$  accordingly.  Nonetheless, we feel  that the
Steidel et al. result provides an observationally-motivated choice and
we adopt the implied normalization for our model constraints in \S5. 

In order to extend our AGN  analysis to $z\gsim3$, we must extrapolate
the   stellar contribution to higher  redshift.   We do so by assuming
that  the stellar emissivity evolves  in the same   manner as the star
formation    rate   density of   the   universe:   $\varepsilon^*(z) =
\varepsilon^*(z=3) \cdot \dot{\rho_*}(z)/\dot{\rho_*}(z=3)$.  For   $z
\gsim 2$, we assume that the star formation rate evolves as 
 
\beq  \dot{\rho_*}(z)  \propto \left(\frac{4z}{1+(z/4.1)^{4.1}}    + 1
\right). \eeq 

\noindent
The shape of this  function provides a  good fit to the star formation
predictions  of Somerville, Primack, \&  Faber (2001) (R.  Somerville,
private communication), and also provides a good representation of the
data, although the scatter in the data is large  (see, e.g. Steidel et
al.~1999,  Poli  et  al.~2001, and the    compilation in figure 12  of
Springel \& Hernquist 2002). 

\subsection{Reemission from Clouds}
\label{sect:reem}

Before going on   to the  results,  we  mention   that another  likely
contributor to  the ionizing flux arises  from the reprocessing by the
IGM of the primary radiation from stars  and AGN (see Haardt and Madau
1996; Fardal  et  al.~1998).   This effectively  transfers   some high
energy photons to lower energies where they are more likely to ionize,
in our case, hydrogen.  The H\,{\sc i} recombinations are right at the
Lyman   edge, so these   photons are   quickly   redshifted to   below
threshold.  However,   He\,{\sc ii}   recombinations  and  two  photon
de-excitations will have     a larger contribution to  the    ionizing
background  (Shull et al.~1999).  Since  stars are not thought to emit
many helium ionizing photons,  this  reemission will mostly  come from
reprocessing   AGN  radiation.   Haardt   \&  Madau  found  that  this
reemission increases the hydrogen ionization rate due to QSOs by about
40\%, independent  of   redshift.    In   this case  we     can  write
$\hat{\varepsilon}^R=0.4\hat{\varepsilon}^{Q}$.   According to  Fardal
et al.~(1998), this  fraction is $\sim25\%$.   For our fiducial models
explored  in  the next   section, we will  assume  that  reemission is
negligible ($\hat{\varepsilon}^R=0$), but   in \S 5.1 we  explore  how
including $40\%$ reemission will affect our derived AGN LFs.

\section{Results}
\label{sect:results}

We will present our constraints under assumption that the LF takes the
form of Eq. \ref{eq:diff_phi}, and thus has four free parameters to be
constrained  at each  redshift: $\gamma_f$,  $\gamma_b$, $\phi_*$, and
$L_*$.   By forcing  the bright-end  to match  the   results of direct
observations we have two constraints: the SDSS slope ($\gamma_b=2.58$)
and normalization (Eqs.  \ref{eq:Phi26_SDSS} and \ref{eq:Phi_def}).  A
third constraint comes from the ionizing background, which effectively
limits  the integrated faint-end     emissivity.   We fix  the   final
parameter  by first  setting the  faint-end  slope at its low-redshift
value, $\gamma_f = 1.58$,  and then we  explore how our results change
for other values of $\gamma_f$ in \S \ref{sect:vary}.  With $\gamma_b$
and $\gamma_f$ fixed, our  emissivity constraints can be expressed  by
simply quoting the  implied   break  luminosity $L_*$  (or  the  break
magnitude $M_B^*$), since the SDSS normalization  fixes $\phi_*$ for a
given $L_*$. 

Table 1   summarizes the ionizing background models   we  have used to
derive our LF's.  In Model A,  we assume that  the total ionizing rate
at high-$z$ is set by that quoted  by M\&M-E (using the flux decrement
distribution  technique).     Model A also    assumes that   starlight
contributes to the background at a level  consistent with what Steidel
et al.~(2001) found, $f_{\rm esc}=0.16$.  Our second example (Model B)
also  assumes the  M\&M-E  ionizing rate   but  now with a  negligible
stellar contribution   $f_{\rm esc} = 0.0$.  Finally,   in Model C, we
assume  the (higher) ionizing rate inferred  from the proximity effect
analyses   with $f_{\rm  esc}   =  0.0$  (we   explore  other  stellar
contributions in \S \ref{sect:Ion_Ss}).  For each model, we obtain our
full  constrained luminosity function  by  iteratively solving for the
value of  $L_*(z)$ that is  consistent with our  adopted ionizing rate
$\hat{\varepsilon}^{Q}(z)$.      Of course     we   change  $\phi_(z)$
accordingly to match the SDSS normalization Eq.~\ref{eq:Phi26_SDSS}). 

\begin{table*}[ttt] 
\begin{center}
\begin{tabular}{||c|c|c|c|c|c|c||}
   \hline          
Model & $\Gamma$  & $f_{esc}$   & $\hat{\varepsilon}^{Q}(z=3,4,5)$
      & $M_B^*(z=3,4,5)$  
      & $N_{HDF}$               & $N_{Keck}$              \\ \hline 
A     & M\&M-E    & 0.16        & 0.74, 0.40, 0.24 
      & -25.8, -24.6, -23.4
      & $0.16^{+0.09}_{-0.07}$  & $79.5^{+37.4}_{-35.8}$ \\ \hline 
B     & M\&M-E    & 0.0         & 1.44, 1.15, 0.86 
      & -24.2, -22.3, -20.8
      & $0.75^{+0.11}_{-0.29}$  & $231^{+92.4}_{-67.0}$ \\ \hline
C     & Proximity & 0.0         & 8.25, 6.92, 2.36 
      & -20.6, -18.7, -18.7
      & $4.18^{+2.09}_{-0.75}$  & $1790^{+322}_{-465}$ \\ \hline
\end{tabular}
\end{center}
Table   1  --  Our  three assumed    (model)  cases for   the ionizing
background.  The first  column lists the  model name, the second lists
the adopted measurement  of  the ionizing rate,  and  the third column
lists  the escape fraction of ionizing  radiation  from stars ($f_{\rm
esc}=0$ implies a  negligible stellar contribution to the background).
The  fourth column gives  the weighted emissivity (Eq.~\ref{eq:eplim})
of AGN  at $z=3,4,5$.  The fifth gives  the break  magnitude ($M_B^*$)
for redshifts 3,  4, 5.  The sixth and  seventh column give the number
of counts that  would have been expected for  the HDF faint AGN search
(\S  \ref{sect:HDF}) and    the Steidel   search  (\S \ref{sect:Keck})
respectively. 
\label{tab:models} 
\end{table*} 

\begin{figure}[t]
\centerline{\epsfxsize=8.5cm \epsfbox{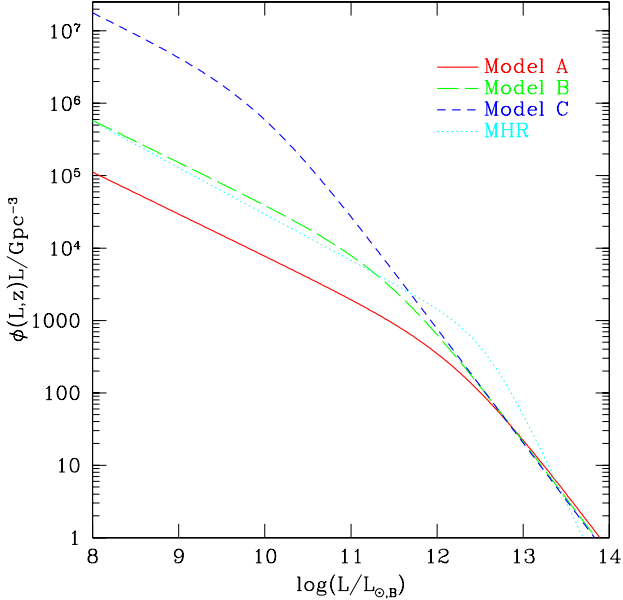}} 
\caption{ The Luminosity Function for our different models at $z=3.5$.
Notice that all three models agree at  the bright-end and only diverge
due to the location of the break.  For comparison, we also plot the LF
from MHR.  }\label{fig:phiL_z3} 
\end{figure}

The change in break luminosity for each model is illustrated in Figure
\ref{fig:phiL_z3},   where we  plot   our  results at   $z=3.5$.  (For
comparison, we also show the MHR LF at  the same redshift.)  Note that
in  the limit that $L_{\rm min}  \ll L_* \ll  L(M_B)$, our constraints
follow the analytic relation 

\beqa \log\left(\frac{L_*(z)}{\rm{L_{\odot,B}}}\right)  &\approx& 12.2
-   1.72  \log   \hat{\varepsilon}^{Q}(z)  -  0.81(z-3)  \nonumber  \\
\log\left(\frac{\phi_*(z)}{\rm{Gpc}^{-3}}\right) &\approx& 2.80 + 2.72
\log \hat{\varepsilon}^{Q}(z) + 0.81(z-3).  \eeqa 

\noindent In practice we use these  expressions as a starting point in
our iterative solution for $L_*$ (and the implied value of $\phi_*$). 

\begin{figure*}[t]
\centerline{\epsfxsize=16cm \epsfbox{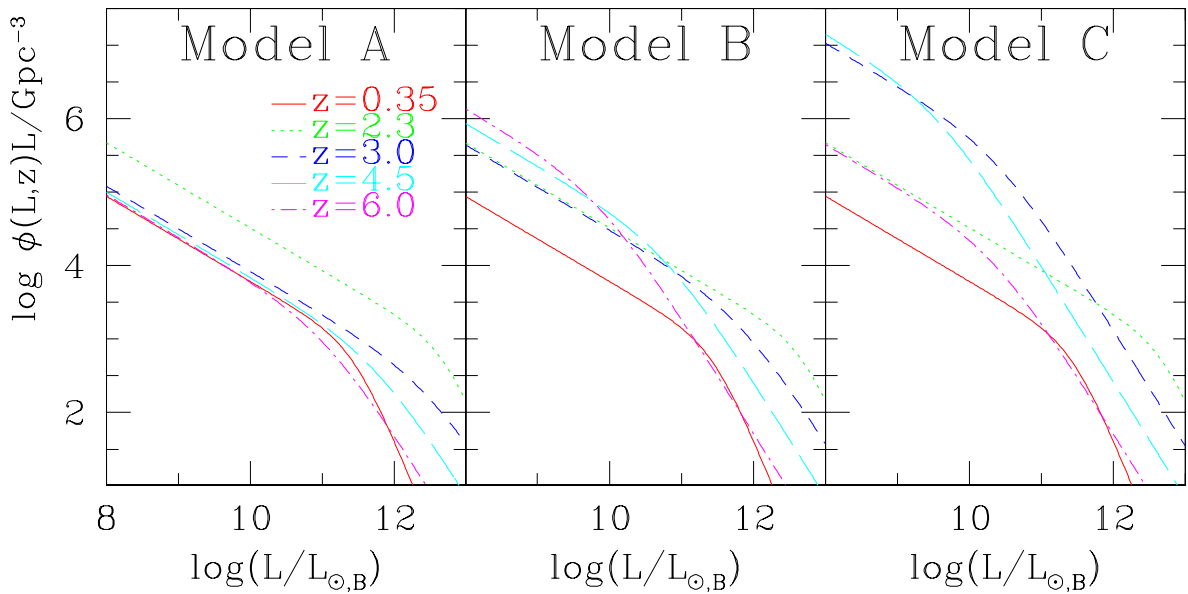}} 
\caption{ The    Luminosity Function for   different redshifts  in our
Models A, B, and C.  The lines for $z=0.35$ and 2.3 are set by the 2dF
data and,        therefore,   are common    to   all     three models.
}\label{fig:3phi} 
\end{figure*}

\begin{figure}[t]
\centerline{\epsfxsize=8.5cm \epsfbox{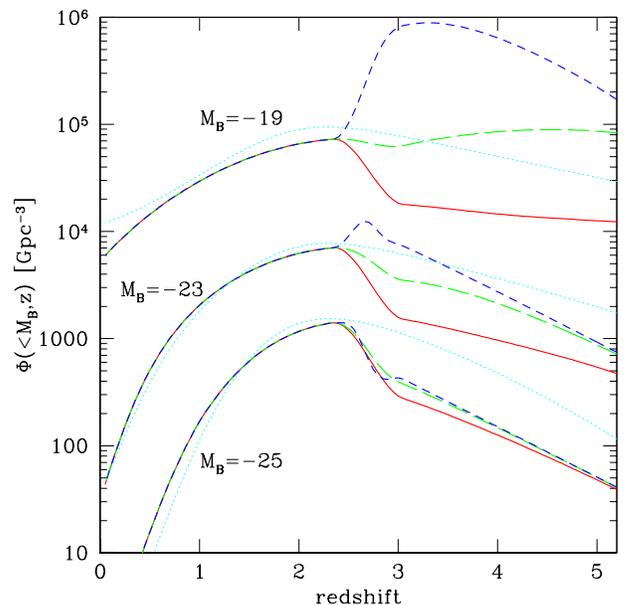}} 
\caption{  The  evolution  of the  integrated   number  counts for the
separate   models,       with  lines  defined     the    same   as  in
Fig.~\ref{fig:phiL_z3}.  }\label{fig:Phi_mod} 
\end{figure}

A direct  presentation of each  Model LF at several discrete redshifts
is  shown  in Figure \ref{fig:3phi}.    For the ionizing background of
Model A, the faint-end of the LF is limited to virtually no evolution,
while Model C allows considerable variation in the faint-end.  The way
in which these  models evolve in  terms of integrated number counts is
illustrated in Fig.  \ref{fig:Phi_mod}: for bright limiting magnitudes
they show very similar   evolution (by construction), while   at faint
magnitudes  the   evolution varies  dramatically.    Note that we have
applied our emissivity constraints  only for $z\geq3$.  For $z\leq2.3$
we assume that the  LF is fully defined  by the 2dF  results.  Between
these redshifts  we have simply interpolated  the values  of $\phi_*$,
$L_*$, and $\gamma_b$.   The awkward line shapes  from $z=2.3$ to 3 in
Fig.  \ref{fig:Phi_mod} are due to this crude interpolation and should
not be regarded as explicit predictions. 

With our Model luminosity functions in hand, we are now in position to
discuss   our results in the context    of several faint AGN searches.
Before going on to do  so, we briefly explore  the degree to which our
results depend on our input parameter assumptions.

\subsection{Variations on Input Parameters}
\label{sect:vary}

\begin{figure}[t]
\centerline{\epsfxsize=8.5cm \epsfbox{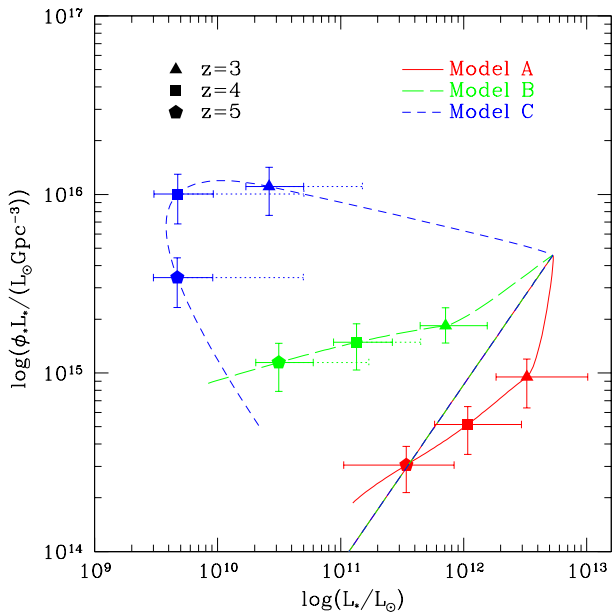}} 
\caption{ Solutions to  the emissivity constrained models for $0<z<6$.
The straight  diagonal line is from the  2dF LF  ($0<z<2.3$), which we
force all of our models to agree with.  The solid error bars come from
varying parameters  in the theory  (see text).  The dotted  line error
bars  are  specifically from  taking the  SSG  value   for $\gamma_b$.
}\label{fig:Lphi_Lz} 
\end{figure}

The evolution of $\phi_*$ and $L_*$ for  our three ionizing background
Models is illustrated  in Figure \ref{fig:Lphi_Lz},  where we show our
results  using  the parameter space   introduced  in conjunction  with
Figure 1 (\S  1).  The solid, dashed,  and short-dashed lines show how
models A, B, and C evolve  in this parameter  space from $z=0$ to $6$,
with triangles, squares, and pentagons  marking specific redshifts: 3,
4,  and 5, respectively.   All of  the models  overlap on the diagonal
line representing the observed 2dF LF evolution  over the range $0 < z
< 2.3$. 

In order to estimate to what degree our input assumptions might affect
our  results,  we have  recalculated our constraints  for  a series of
cases   in which  each of  our  inputs   was  varied (while  all other
parameters were  kept fixed).  The  changes we explored  were: varying
$\gamma_f$  by $\pm 0.2$;  varying  $\eta_\nu$ by $\pm 40\%$;  letting
$\gamma_b  =  2.87$ (rather  than our   fiducial  $\gamma_b =  2.58$);
allowing  for non-zero cloud  reemission of $\hat{\varepsilon}^R = 0.4
\hat{\varepsilon}^Q$  (see \S  \ref{sect:reem}); and  incorporating an
IGM model   with  more absorption  at high-redshift   (as advocated by
Fardal et  al.~(1998), see Appendix).  We used  this suite of cases to
estimate plausible uncertainties in  our  model LFs at each  redshift,
and  utilize these uncertainties in  the  next subsections.  The error
bars  on the different  redshift   points in Figure  \ref{fig:Lphi_Lz}
represent the  largest changes in   $\phi_*$ and $L_*$ we observed  at
each redshift for  the changes described above.  As  can  be seen from
the   figure, the   primary changes are   due  to  the adopted   Model
backgrounds,  with    variations  on  input  parameters   giving  less
important, although noticeable    changes  in the   characteristic  LF
scales. 

Probably the  most straightforward change  we observe  comes from from
varying $\eta_\nu$.  Increasing (decreasing) $\eta_\nu$ requires fewer
(more) faint AGN to reproduce the same ionizing background.  Obtaining
fewer AGN requires an increase in  $L_*$ with a corresponding decrease
in $\phi_*$ because the   two parameters are  constrained by  the SDSS
normalization:     $\phi_*    \propto     L_*^{1-\gamma_b}$    (recall
Fig.~\ref{fig:phiL_Lz4}).  For our $\pm 40\%$ variation in $\eta_\nu$,
the effect on $\phi_*$ is about a factor of 2  , while $L_*$ varies by
about 50\%.  Adding IGM reemission acts in  the same way as increasing
the  value   of $\eta_\nu$.    In   going    from no  reemission    to
$\hat{\varepsilon}^R =0.4  \hat{\varepsilon}^Q$,  $L_*$ increases by a
factor of $\sim2$,  and  $\phi_*$ decreases by  a  factor  of $\sim3$.
Employing  a model with more  attenuation at high  redshift (Fardal et
al.~1998, Appendix) requires more AGN  to match a given ionizing rate.
The effect is most prominent in Model A for $z\gtrsim4$.  It increases
$\phi_*$ by nearly a factor of 6. 

\begin{table*}[ttt] 
\begin{center}
\begin{tabular}{||c|c|c|c|c||}
   \hline          
Model & $\gamma_f$  
      & $M_B^*(z=3,4,5)$  
      & $N_{HDF}$               & $N_{Keck}$              \\ \hline
A     & 1.38
      & -25.2, -24.1, -22.9
      & 0.15 & 66.0 \\ \cline{2-5}
      & 1.78
      & -27.0, -25.7, -24.3
      & 0.17 & 90.6 \\ \hline
B     & 1.38
      & -23.7, -21.9, -20.4
      & 0.79 & 219 \\ \cline{2-5}
      & 1.78
      & -25.2, -23.1, -21.4
      & 0.67 & 231  \\ \hline
C     & 1.38
      & -20.3, -18.4, -18.4
      & 4.64 & 1990 \\ \cline{2-5}
      & 1.78
      & -21.2, -19.2, -19.2
      & 3.55 & 1520 \\ \hline
\end{tabular}
\end{center}
Table 2 -- We recalculate each model with  a higher and lower value of
the faint-end slope.  In comparison to our fiducial values in Table 1,
the changes to  the   break magnitude  are  well-approximated  by  Eq.
\ref{eq:del_MB}.   Notice,   however,  that   the faint   counts  (see
\S\ref{sect:HDF} and \S  \ref{sect:Keck}) are not  highly-dependent on
the choice of $\gamma_f$. 
\label{tab:gamma_f} 
\end{table*} 

We  have no observational  constraints on $\gamma_f$ at high redshift,
so variations in this parameter are certainly important to explore.  A
simple approximation of  the emissivity integral,  along with the SDSS
normalization,     gives   us   the   variation    to   our   fiducial
($\gamma_f=1.58$) break magnitude: 

\beq              M_B^*(\gamma_f)-M_B^*(1.58)                  \approx
1.6-4.3\log\left(\frac{2.58-\gamma_f}{2-\gamma_f}\right). 
\label{eq:del_MB}
\eeq 

\noindent  By iterating over  the  relevant equations, we obtain  more
precise estimates, which we list in Table  2 for $\gamma_f=1.38, 1.78$
in each  of our models.  A flatter  (steeper) faint-end slope requires
more   (less) AGN   at  ``medium''   luminosities to  match   a  given
background.  This  alteration  is one  of the  dominant variations  on
Model A, since the break here is very near to the limits from the SDSS
constraint (Eq.   \ref{eq:Phi26_SDSS}), and thereby the corrections to
Eq.  \ref{eq:del_MB}  are somewhat larger.   In  fact, at  $z\sim3$, a
steeper $\gamma_f$  would make $M_B^*<-26$.   A break at  such a large
absolute  magnitude would presumably  have been  detected by WHO94 and
SSG.  Indeed, if current indications are correct and $M_B^*>-26$, then
our fiducial Model A is close to  being a {\textit{minimum}} allowable
LF.     This    possibility      will    be examined      further   in
\S\ref{sect:Implics}. 

One last parameter  that we give special  attention  to is $\gamma_b$.
The SDSS  value  of  2.58 is quite  a  bit  smaller than   in previous
determinations.  Similar to the case of $\gamma_f$ for Model A, taking
the steeper value  of 2.87 from SSG has  a relatively large  effect on
Models   B and C  (recall  Fig.~\ref{fig:phiL_Lz4}).   In Model C  the
effect is  nearly an order  of magnitude  in $L_*$.   We plot this  in
Fig.~\ref{fig:Lphi_Lz} with a separate (dotted) error bar, although it
is only visible above the (solid) error bar  in cases where it exceeds
all other uncertainties.  Note that since the  median redshifts of SSG
and   SDSS are, respectively, $\sim3.3$  and  $\sim4$,  it may be that
$\gamma_b$ is evolving over these epochs. 

\subsection{Faint Surveys}

There are several observations (either planned  or completed) that are
relevant to faint magnitude AGN.  The counts  obtained for the deeper,
smaller fields that  have been done so  far  are not as  statistically
significant as   those based on  large,   shallow surveys  we used  to
normalize  our Model  LF's at the  bright-end.   However these fainter
surveys do provide   a useful consistency   test for our   Models, and
forthcoming  surveys  for faint  AGN will  offer  further tests of our
input assumptions  about the ionizing background.   In this section we
briefly investigate several surveys in light of our Model expectations
and also predict what AGN counts might be seen by future observational
programs. 

\subsubsection{Hubble Deep Field}
\label{sect:HDF}

\begin{figure}[t]
\centerline{\epsfxsize=8.5cm \epsfbox{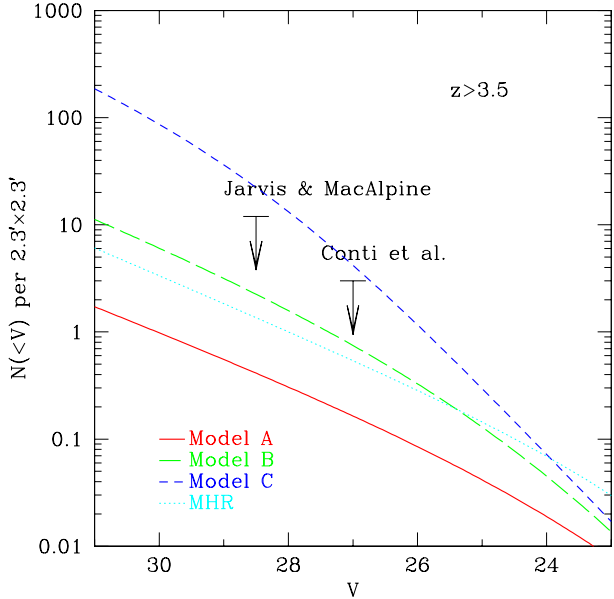}} 
\caption{ Number counts  in the Hubble Deep Field  for $z>3.5$.  Shown
are the predictions from our three models, as well as the upper limits
from two searches for QSO candidates in the HDF.  }\label{fig:HDF} 
\end{figure}

The  Hubble Deep Field  is  approximately 2.3  by 2.3 $\rm{arcmin}^2$.
Looking   only at unresolved sources,   Conti et al.~(1999) claim that
there  are no $z>3.5$   QSO candidates in the   HDF down to  estimated
completeness        limits      of    $\rm{B}_{450}\simeq       27,0$,
$\rm{V}_{606}\simeq27.0$,   $\rm{I}_{814}\simeq26.0$.\footnote{Besides
the attenuation from continuum  absorption, the observed spectrum of a
distant QSO will be affected by line  blanketing in the IGM, for which
we employ the  approximations from  Madau  (1995).  The resulting  QSO
colors   match qualitatively the    color-color  plots from  Conti  et
al.~(1999),   Haiman, Madau, \& Loeb (1999),   and Jarvis \& MacAlpine
(1998).}  The Conti  et al.~results are  consistent with a similar HDF
search by Elson, Santiago, \& Gilmore (1996).  Conservatively, we will
assume that this implies an upper  bound of 3  AGN per HDF field.  For
Poisson statistics, this   corresponds to $5\%$  probability of seeing
zero AGN.  At slightly  fainter magnitudes, Jarvis \& MacAlpine (1998)
take resolved sources in  the HDF and  select those with nuclei having
the     expected     AGN  colors.        In the     magnitude    range
$\rm{V}_{606}\sim27.0-28.5$, for $z>3.5$, they find 12 candidates, but
they make it  clear  this  is   an  upper limit.   We  illustrate  the
observational bounds along with our three model expectations in Figure
\ref{fig:HDF}.    Interestingly,   there    is   one   high   redshift
($z\approx3.5$)    QSO seen  by   Chandra     in  the HDF     (CXOHDFN
J123639.5+621230.2,  see    Hornschemeier  et  al.~2001;    Brandt  et
al.~2001).  This object is not selected by  either Jarvis \& MacAlpine
(1998) or Conti et al.~(1999), even though it has $R=24.3$. 

Using the Conti et  al.~limits ($\rm{V}_{606}<27.0$, $z>3.5$), we have
calculated what each  of our Models  would  have expected in  the same
field.  These values are listed under $N_{HDF}$ in Table 1, along with
an  uncertainty calculated by the  same  variation of input parameters
discussed  in the previous section.  Model   C predicts too many QSOs.
Even allowing  $\gamma_b$ to be  flatter  than the SDSS  value  cannot
reduce the expected  value below $\sim 3$  per field.   We discuss the
implications of this in \S 6. 

\subsubsection{Keck QSOs}
\label{sect:Keck}

Unlike most other surveys, which have targeted QSOs by their strong UV
emission, Steidel and collaborators have only taken spectra of objects
with  strong  continuum breaks at the   rest-frame Lyman limit.  Their
technique  was designed   to  find high-$z$ galaxies,   which  have an
intrinsically softer spectrum  at  the Lyman continuum.  Type   I QSOs
have much harder continua, so they would be  generically missed by the
Lyman-break selection technique.   However an estimated  $\sim60\%$ of
the  lines of sight  to $z\sim3$  QSOs should have  a foreground Lyman
limit system at  high   enough redshift to create   an  ``artificial''
break.   These  are  the   AGN that   get  targeted for  spectroscopic
identification.  Steidel  et al.~(2002)   identified 13 type   I  QSOs
between redshifts of 2.7 and 3.3 down to $\mathcal{R}=25.5$.  Based on
the size of their field and  a rough estimate of their incompleteness,
the implied number  density of type I  AGN is  approximately $190$ per
square degree.\footnote{ Steidel et al.~(2002) state that more careful
analysis will be done in a later paper.} 

We have compiled the expected number per square degree  in each of our
Models in Table  1 (with uncertainties determined  as before).  Again,
Model C  seems to over-predict  (dramatically) the  number of high-$z$
QSOs, while   Model B,  is   consistent, at  least   within reasonable
uncertainty.  Interestingly, the  LF motivated by the  Steidel group's
own estimate for the stellar escape fraction  (Model A) seems to under
predict the  expected QSO  abundance  by a  factor of  $\sim  2$.   We
discuss this in the next section. 

Also detected in  the same survey were 16  narrow lined, type II QSOs.
These  locally absorbed objects are quite  important to the X-ray and,
perhaps, IR  backgrounds, but they  are  likely very  weak emitters at
ionizing frequencies,  and  therefore do  not  directly relate to  the
constraints we have set  up.  We will  also address this population in
\S 6.

\subsubsection{Future Surveys}
\label{sect:FutureSurveys}

There are a number of future and ongoing  surveys that plan to fill in
the faint high redshift portion  of the AGN LF.   The Big Faint Quasar
Survey (Hall 2001) will presumably probe the LF for $-26.5<M_B<-23.5$,
for  $z>3.5$.   This is  below the   luminosity range  of  SDSS.  Hall
predicts that this survey (for  $R<23.5$ and $7\rm{deg}^2$) will  find
$\sim120$ quasars for  $3.5<z<4$,   and $\sim40$ for   $4<z<5$.  While
Model A with a  substantial stellar  contribution would predict  fewer
detections  than this, Models  B  and C with AGN-dominated backgrounds
expect 2-3 times as many QSOs. 

\begin{figure}[t]
\centerline{\epsfxsize=8.5cm \epsfbox{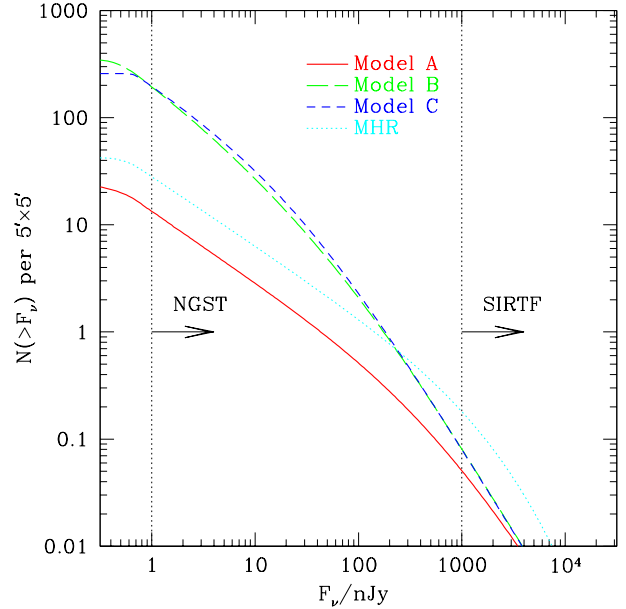}} 
\caption{Predicted  number  counts in   the  $1-3.5\mu\rm{m}$ band for
$z>5$  for our three  models, as well  as for MHR.  The vertical lines
are  the   proposed     flux limits   for  future  space   telescopes.
}\label{fig:NGST} 
\end{figure}

The Space Infra-Red   Telescope Facility (SIRTF)  is proposed  to have
1$\mu$Jy sensitivity, and the  Next Generation Space Telescope  (NGST)
is   proposed  to  have  1nJy sensitivity  between    1 and 3.5$\mu$m.
Although selecting  high  redshift QSOs in the  Infra-Red  may not  be
straightforward (see Warren \& Hewett  2002), it is worthwhile to work
out our theoretical predictions at  these detection thresholds.  These
are shown in Figure \ref{fig:NGST}. 

Since all of our Models drop off steadily for $z>5$, the number counts
illustrated in Figure  \ref{fig:NGST} are  dominated by the   $z\sim5$
QSOs.  At this redshift for our chosen SED and cosmology, the proposed
flux limits of  SIRTF (NGST)  would  correspond to AGN  with $L \simeq
10^{11} L_{\odot}$ ($10^8 L_{\odot}$).  We can  compare our results to
those  of  Haiman  \& Loeb   (1998),  who used  a model   based on CDM
structure formation to  predict that SIRTF would  observe on the order
of a few  $z > 5$ AGN, while  NGST would see approximately $10^{3.5}$.
A close examination of Haiman \& Loeb's differential LF (their Fig. 5)
at  $L \simeq  10^{11} L_{\odot,B}$   shows that it    is an order  of
magnitude higher than the  simplest extrapolation of the SDSS results.
This explains  why  their SIRTF prediction  is more  than an  order of
magnitude higher  than  any of our predictions.    So  too, their NGST
estimates are a factor of ten to a hundred above any of our estimates.
However,  this  appears to be   due  to the  fact   that  their LF  is
practically   constant    for       $L   <10^{10}   L_{\odot}$     and
$z>5$.\footnote{Note that  Haiman \& Loeb  have since  re-tooled their
model (Haiman,  Madau \&  Loeb 1999; Haiman   \& Loeb  1999),  further
suppressing the  formation of small  mass black holes, and the updated
NGST predictions appear  to be more in line  with our Models B and  C.
Still, however, the predicted shape of the \emph{bright} end of the LF
from a simple scaling  of the CDM mass  function would appear to be in
conflict   with the SDSS   observations,   and therefore,  it  may  be
necessary to  employ some  sort of feedback  mechanism to  curtail the
formation of intermediate AGN.}

\section{Implications}
\label{sect:Implics}

In addition to  direct constraints on the nature  of the faint-AGN LF,
our  results have  some interesting   implications  for the  nature of
ionizing sources  as well  as  some   typical characteristics of   AGN
themselves. 

\subsection{Ionizing Sources}
\label{sect:Ion_Ss}

Our joint analysis of the ionizing rate and  faint AGN counts provides
a  potentially useful avenue for exploring  the  contribution of stars
and AGN to   the ionizing background.   We  showed in \S 5  that faint
surveys do not find the  number of QSOs that would  be expected if all
of the proximity-effect-derived   background is coming   only from AGN
(Model  C).  Conversely, a model that  reproduces the (lower) ionizing
background  intensity   favored   by  the  flux    decrement technique
under-predicts  the faint  counts  unless the stellar contribution  is
limited   (Model B).  In this  subsection   we attempt  to place  more
quantitative limits on the   stellar contribution implied by the   two
competing background intensity measurements. 

\begin{figure}[t]
\centerline{\epsfxsize=8.5cm \epsfbox{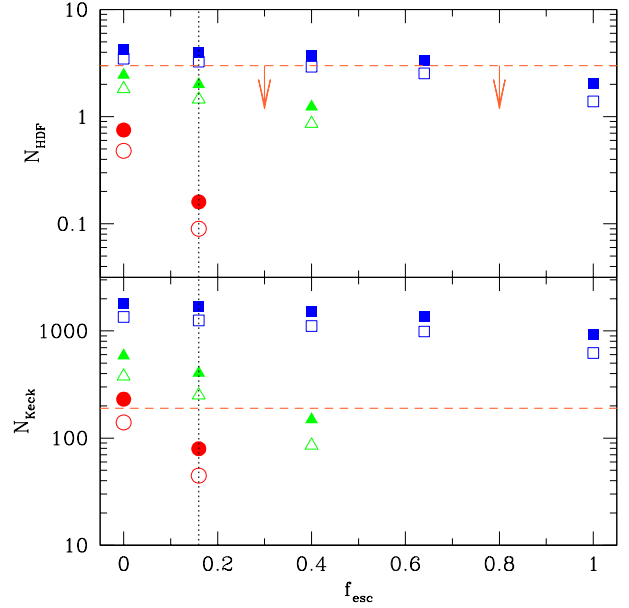}} 
\caption{ The expected number counts in  the HDF (\emph{top}) and Keck
(\emph{bottom}) surveys  as a  function of  the escape  fraction.  The
upper bound   on  the HDF  and   the observed number from   Steidel et
al.~(2002)  are plotted as  dashed horizontal  lines.  The squares are
for an ionizing background from the proximity effect (as parameterized
by Fardal et al.~1998); the triangles are from  the lower bound on the
proximity  effect from Scott et al.~(2000);  and the  circles are from
the flux  decrement measurement of M\&M-E.   The open/solid points are
with/without    the         contribution  from      cloud   reemission
($\hat{\varepsilon}^R =  0.4\hat{\varepsilon}^{Q}$).  Note that  we do
not plot points at  high escape fraction for  the flux decrement  (and
for the  lower bound  on  the proximity effect,   as well),  since the
resulting    LF   breaks    would    contradict   observations    (see
\S\ref{sect:vary}).  }\label{fig:f_esc} 
\end{figure}

For each  ionizing   background measurement,  we  have calculated  the
expected number of  Keck  and HDF  counts for  several  values  of the
escape fraction (\S\ref{sect:stars}).  The results are shown in Figure
\ref{fig:f_esc}, with the different  symbol types reflecting different
background assumptions: the  squares  represent a standard   proximity
effect background (Fardal et al.  1998) and  the circles correspond to
an assumed flux decrement background.  In order to be conservative, we
also include results for a background  that matches the lower limit on
the proximity effect rate taken  from Scott et al. (2002) (triangles),
with    $\Gamma_{-12}\simeq1.0$      over  the    range   $1.7\lesssim
z\lesssim4.1$.  Solid  symbols   include no  IGM  reemission and  open
symbols include $40\%$ cloud  reemission.  The horizontal dashed lines
indicate   the  upper limit from  the  HDF  search  (upper  panel) and
observed AGN per square degree seen in the Keck fields (lower panel). 

By   examining Figure  \ref{fig:f_esc}  we see   that  even with  (the
 unphysically large value)  $f_{\rm  esc}=1$, the  standard  proximity
 effect background yields far too many Keck QSOs.  A contribution from
 IGM reemission  (\S \ref{sect:reem})    does little to    lessen  the
 discrepancy.  The lower  bound on the proximity  effect from Scott et
 al. provides more reasonable  numbers, but does require a significant
 stellar component,  $f_{\rm esc} \gsim 20\%$, even  if we allow for a
 large  amount   of reemission.\footnote{   A  similar  discussion  on
 acceptable escape fractions in light of proximity effect measurements
 can be found in Bianchi et al.  (2001), but there they assume PLE for
 their AGN  LF, similar to  that of MHR. }  The flux-decrement-derived
 background, on the other  hand, is not  compatible with much  stellar
 contribution:  $f_{\rm  esc}  \lsim  5\%$  is  required   in order to
 reproduce   the Keck counts.\footnote{We  stress   that  these escape
 fraction  limits cannot be more  precise, as the uncertainties in the
 Keck survey are not yet known.} 

An additional way  to examine the question  of stellar escape fraction
involves  using the He Ly$\alpha$ forest  to limit the relative amount
of ionizing radiation from  stars and AGN.   Stars, unlike AGN, do not
have much emission  that extends to the helium  Lyman continuum.  This
means that a    stellar-dominated background  should have a    smaller
relative fraction of fully ionized helium.  At $z\sim2.5$, the implied
optical   depth to He\,{\sc   ii} absorption  is  low enough  that the
background at these redshifts should be QSO-dominated (Davidson et al.
1996).  But at slightly  higher redshifts, Heap  et al.  (2000) appear
to detect a Gunn-Peterson trough  blueward of helium Ly$\alpha$ in the
spectra of Q0302-003.   They conclude that   the ratio of  hydrogen to
helium    ionization rates    rises     abruptly at   $z\approx3$   to
$\Gamma_{\rm{HI}}/\Gamma_{\rm{HeII}}\simeq800$, suggesting that a soft
stellar  contribution  is  beginning  to  dominate the  hard  ionizing
spectrum  of AGN at  this epoch.  This kind  of rapid hardening of the
background at $z\sim3$ is supported by the analysis of Songaila (1998)
of the Si\,{\sc iv}/C\,{\sc iv} ratio in the IGM.   Although we do not
model  the  propagation   of  helium  ionizing  photons   in   a  very
sophisticated    way     (see  Appendix),     we    show    in  Figure
\ref{fig:Gamma_Gamma} that Model A  reproduces the reported  evolution
of $\Gamma_{\rm{HI}}/\Gamma_{\rm{HeII}}$ in a qualitative way.  A more
careful analysis of  this phenomena  will  likely have to include  the
concomitant  reionization of He\,{\sc  ii} (see Theuns  et al.  2001).
Recently,  Sokasian, Abel,  \&.   Hernquist (2002)  have used observed
opacities of   HI  and HeII  to  conclude   that stars and    AGN must
contribute  roughly equally to the ionizing  background at $z \sim 3$.
Theuns  et al.  (2002)  and  Bernardi et al.   (2002)  argue  that the
evolution  of the flux decrement  distribution studied via a sample of
SDSS  quasars   points   to    HeII  reionization   at    $z    \simeq
3-4$.\footnote{In writing this   paper, we explored another   idea for
separating the relative  galaxy/AGN  contribution  to  the background.
This  was to use $\sim \rm{GeV}$  gamma ray attenuation to measure the
UVB as  is done for the  IR  background using $\sim  \rm{TeV}$ sources
(see  Primack et al.  2001,  Bullock et al.  2002).  Unfortunately, at
these redshifts, UV  photons are completely  overwhelmed by foreground
optical photons, because the peak energy of  the interaction scales as
$(1+z)^{-2}$.  One  can phrase this  result in the positive: gamma ray
attenuation is sensitive only to the integrated stellar light over the
history of the universe.  AGN consistent with current estimates of the
ionizing background   will provide a   negligible contribution  to the
extra-galactic background light relevant for gamma-ray attenuation.} 

\begin{figure}[t]
\centerline{\epsfxsize=8.5cm \epsfbox{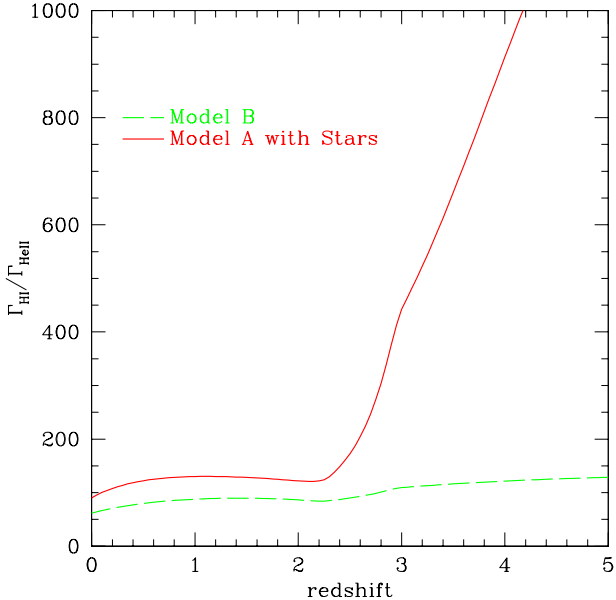}} 
\caption{                  The            softness          parameter,
$\Gamma_{\rm{HI}}/\Gamma_{\rm{HeII}}$, for Model   B (QSOs  only)  and
Model A (stars and QSOs).  The rapid rise at $z\sim 3$  for Model A is
qualitatively  similar to  that  reported   by  Heap et al.    (2000).
}\label{fig:Gamma_Gamma} 
\end{figure}

\subsection{AGN Characteristics}
\label{AGN_BH_densities}

In  addition to  quantifying our expectations  for AGN  counts in  the
presence of  different   ionizing   backgrounds,   our  results   have
implications for the emission efficiency and lifetimes of AGN. 

As discussed in \S 1, AGN activity is  likely driven by accretion onto
super-massive black  holes.   Under this  assumption, the  present-day
density of  black holes  is related  to the integrated  AGN emissivity
over the  history  of the  universe,  modulo the AGN  efficiency  (see
Soltan 1982;   Chokshi  \& Turner  1992;  Haehnelt, Natarajan  \& Rees
1998).  Because our models   provide   estimates of the maximum    AGN
emissivity out to high redshifts, we can  convolve our results with an
estimate of the density of black holes today in  order to determine an
implied efficiency. 

Specifically the relic black  hole density $\rho_{\bullet}$ is related
to the  integrated    QSO  energy density  by   $\rho_{Q}  =  \epsilon
\rho_{\bullet}$,  where $\epsilon\equiv L_{bol}/\dot{M}c^2$ is the QSO
efficiency of converting mass into energy.  The total energy output of
QSOs can be obtained by integrating the emissivity over the history of
the       universe.                  If               we        define
$\varepsilon_B(z)=\nu_B\varepsilon_B(\nu_B,z)$ and  $\nu_B=c/4400$\AA,
then we can write 
 
\beq 
\label{eq:rhoq}
  \rho_{Q} =  \frac{L_{bol}}{\nu_B  L_\nu(\nu_B)} \int \frac{dt}{dz}dz
	       \cdot\varepsilon_B(z).  \eeq 

\begin{figure}[t]
\centerline{\epsfxsize=8.5cm \epsfbox{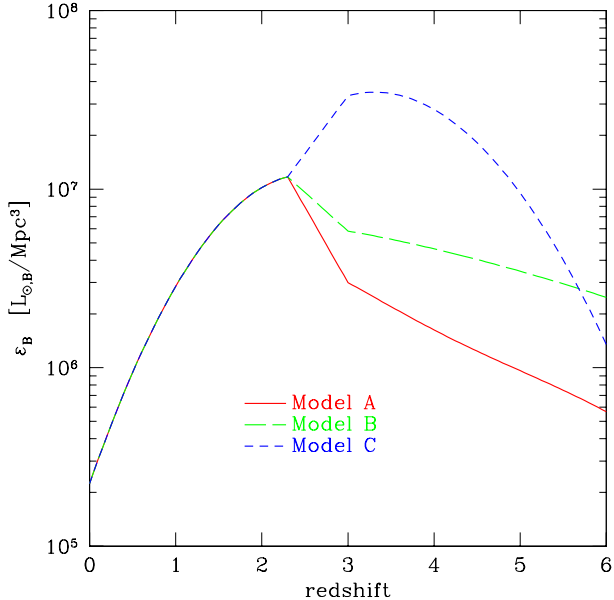}} 
\caption{ The  emissivity  in the B  band.  At  low redshift we simply
take the fit from  2dF.  For $z>3$, we plot   our three models with  a
simple linear  interpolation to the  2dF  limit of $z=2.3$.   We  have
adopted units  such  that   one   $\rm{L_{\odot,B}}$  corresponds   to
$2.11\times10^{33}\rm{erg/s}$ (see SSG).  }\label{fig:eps_B} 
\end{figure}

\noindent 
Because we are interested in the total  energy output, we have applied
a bolometric correction:  $L_{bol}/\nu_B  L_\nu(\nu_B)=11.8$ (Elvis et
al.   1994).  We plot  the B band  emissivity  for our three models in
Figure \ref{fig:eps_B}.  For $z<2.3$ we have used  the 2dF LF fit from
Boyle et al.  (2000) to determine  the emissivity, and between $z=2.3$
and 3, we linearly  interpolate from the 2dF fit  to our high redshift
constraints.   When we integrate from $z=0$  to 6, we find that Models
A, B, and C give $\rho_{Q} = 1.38,  1.57,$ and $2.81$ respectively, in
units  of $10^4 \rm{M_\odot Mpc^{-3}}$.  The  low-$z$ LF  from the 2dF
alone gives $\rho_{Q}(z\le  2.3) = 1.12$   in the same units.   Thus a
large fraction of the energy output from AGN seems to have occurred at
late times even for our most extreme Model (C), and more than half for
Models A and B . 

The present day density in black holes  can be estimated by assuming a
typical black hole mass to  spheroid mass ratio, $M_{\bullet}/M_{sp}$,
and  combining it with a local  determination for the mass fraction in
spheroids $\Omega_{sp}$.    Adopting     the    fiducial   value    of
$M_{\bullet}/M_{sp}  = 0.13\%$  (Merritt \& Ferrarese  2001; McLure \&
Dunlop  2002;   van der Marel  1999)   and  $\Omega_{sp}=0.002 h^{-1}$
(Fukugita et al. 1998), we obtain 
 
\beq       \rho_{\bullet}           =     5.0        \times       10^5
\left(\frac{M_{\bullet}/M_{sp}}{0.0013}\right)
\left(\frac{\Omega_{sp}}{0.002h^{-1}}\right)
\left(\frac{h}{0.7}\right)\rm{M_\odot Mpc^{-3}}. 
\label{eq:rho_BH}
\eeq 

\noindent 
Salucci   et     al.       (1999)   obtained    a     similar    value
($8.2h^2\times10^5\rm{M_\odot Mpc^{-3}}$) by convolving a distribution
of  $M_{\bullet}/M_{sp}$ values with an  estimate of the spheroid mass
function (see  also Yu \&  Tremaine 2002).  Adopting $\rho_{\bullet} =
5.0\times10^5   \rm{M_\odot   Mpc^{-3}}$,  we  obtain   $\epsilon    =
\rho_{Q}/\rho_{\bullet} = 0.028, 0.032,$  and $0.056$ for models A, B,
and C respectively. 

Note that although these are in  principle maximum efficiencies, based
on  maximum  allowable  AGN   emissivities (because  we  have  ignored
reemission), they   are significantly  smaller than the  often-adopted
value  of $\sim 0.1$.     In  addition, our derived  efficiencies  are
significantly   smaller  than those   obtained   using the hard  X-ray
emissivity (Fabian \& Iwasawa 1999; Salucci et al.  1999; Elvis et al.
2002).  One possible explanation for this discrepancy is the existence
of  a   large population of obscured   AGN.   If AGN  efficiencies are
typically $\gsim 10\%$  (e.g.  Elvis et  al.  2002),  then our results
require that $\gsim ~ 50 - 70\%$ of  AGN are significantly obscured in
the  UV and optical.   Interestingly, this  result  is similar to that
obtained by synthesis modelers,  $\sim75-90\%$, based on the shape  of
the   X-ray background  spectrum (Comastri   et  al.   1995; Gilli  et
al. 2000) . 

We have not  considered type 2 AGN  in our analysis simply because the
implied  absorption would mean  they are  not strong ionizing sources.
But unification  models assume  that   the type  2 population  is  not
separate from the type  1s, but  merely  the result of  an orientation
effect.  This would imply    a non-trivial mapping between   the  mass
function of   supermassive  black  holes and the    optical luminosity
function.  Therefore, to study the  accretion history of the universe,
it would seem to be more straightforward to use bands less affected by
obscuration  (see Marconi \&   Salvati   2001; Barger  et   al.~2001).
Unfortunately, the X-ray LF at high  redshift is limited statistically
(Miyaji et  al.  2000),  and the AGN  activity in  many IR sources  is
still a matter of debate  (see Lawrence 2001; Fadda  et al.~2002).  At
high redshift, the best  limits on the shape  and evolution of the AGN
LF still come from the optical/UV. 

Another constraint related to the  local number density of black holes
concerns the fraction of supermassive  black holes that are active  at
any time, $f_{\rm   on}(z)$.  This can  be  related to a typical   AGN
lifetime,  $t_{agn}$, which     we  define,  via $f_{\rm   on}(z)    =
t_{agn}(z)/t_H(z)$, to be the time that a typical black hole is active
over the course of a Hubble time.  For  simplicity, let us assume that
every AGN  shines at  a fixed fraction  of the   Eddington luminosity:
$L=L_\lambda\equiv\lambda             L_{Edd}$,                  where
$L_{Edd}\approx6\times10^{3}(M_{\bullet}/\rm{M_\odot})
\rm{L_{\odot,B}}$.   With   this  assumption, we  write   $f_{\rm on}$
(averaged over luminosity) as 

\beq f_{\rm on}(z) \equiv \frac{\Phi(>L_\lambda,z)}{N(>M_{\bullet},z)}
  \geq \frac{\Phi(>L_\lambda,z)}{N(>M_{\bullet},z=0)}. 
\label{eq:f_on}
\eeq 
 
\noindent 
The inequality comes  from the  assumption that  a given black  hole's
mass only grows with time and  that the black hole  mass function is a
decreasing function of mass.   One possible caveat to  this inequality
is that  black hole  merging could  (but  not necessarily) reduce  the
number  of low-mass black  holes with time.   In  this case, the above
inequality  would break down.  However, in  the  case of halos of mass
$\sim  10^{10}  M_{\odot}$ (corresponding  to  black  holes   of $\sim
10^{6}M_{\odot}$), the  number density does not decrease substantially
from $z \sim  5$ to the present.   This is because most merging occurs
in high-mass-ratio events. 

\begin{figure}[t]
\centerline{\epsfxsize=8.5cm \epsfbox{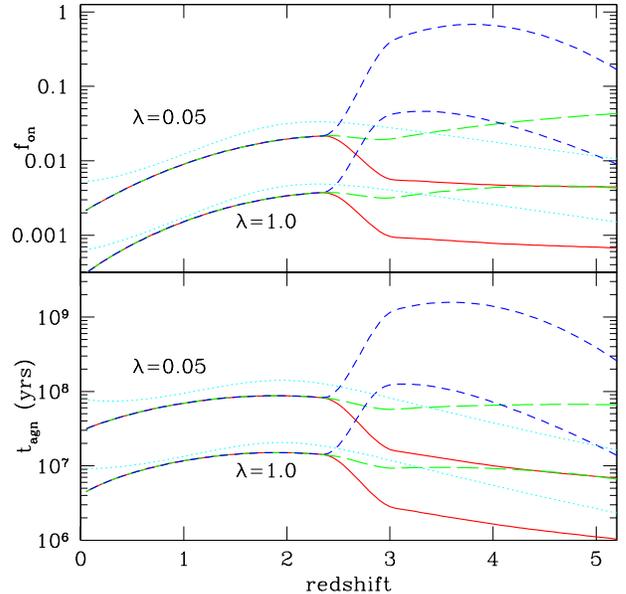}} 
\caption{\emph{Top}-  lower bounds  on   the fraction of   $10^{6}{\rm
M_\odot}$ black holes that  are active at any redshift; \emph{bottom}-
implied  lower bounds  on typical   AGN  lifetimes  (see  text  for  a
description).    We  plot two   assumptions   for  the  ratio  of  AGN
luminosities to  the  Eddington luminosity:  $\lambda\equiv L/L_{Edd}$
equal to 0.05 (upper set of curves) and 1.0 (lower).  Our three models
are represented,  along with  the  model of  MHR for  comparison.  The
lines are defined as in Figure \ref{fig:phiL_z3}.  }\label{fig:t_agn} 
\end{figure}

So  assuming Expression \ref{eq:f_on}  is valid,  we  can take for the
denominator the  results of Salucci et  al.  (1999), who estimate that
$N(>M_{\bullet},z=0)      \simeq        10^{-2}\rm{Mpc}^{-3}$      for
$M_{\bullet}=10^6{\rm  M_\odot}$.  Given $\lambda$, this number allows
us  to determine the implied lower  limits on $f_{\rm on}(z)$ for each
of  our Models.  We  plot these  limits  in Figure \ref{fig:t_agn} for
$\lambda=1.0$      ($L_\lambda=6\times10^{9}\rm{L_{\odot,B}}$)     and
$\lambda=0.05$            ($L_\lambda=3\times10^{8}\rm{L_{\odot,B}}$).
Fortunately, none  of  our  models result   in the  unphysical $f_{\rm
on}>1$.  But for the  corresponding lifetime limits, Model C  requires
very long-lived AGN  at  high-$z$: $t_{agn} \gsim  10^{8}$yr,  whereas
Models A and B suggest lifetimes that are  on the high-side of what is
typically  assumed: $t_{agn} \gsim 10^7$yr.    If estimates of the AGN
lifetime from other lines of reasoning can  be obtained (e.g.  Martini
\& Weinberg 2001; Haiman \& Hui 2001), then this  sort of analysis may
be  capable  of restricting theories   on the  history of  black  hole
accretion (see Ciotti et al. 2001).

\section{Summary and Conclusions}
\label{sect:SandC}

In this paper, we have explored the implied evolution of the faint AGN
luminosity  function using three   different  models for  the ionizing
background from $3 \lsim z \lsim 6$.  Our  LFs are derived by matching
the implied faint-end  emissivity.  Unfortunately,  the value of   the
ionizing background rate is not universally agreed upon.  Measurements
obtained using the proximity  effect generally give larger values than
those based on the flux decrement distribution. 

Although the current data on faint AGN  come from small-field searches
with  only modest statistical  accuracy, we have   used them to obtain
rough evaluations  of our Models.  If AGN  were producing a background
at  the level  measured by   the   proximity effect technique,    then
significantly more would  have been seen  in  the HDF and LBG  fields.
For typical proximity effect values  of the background, even an escape
fraction of unity would require  more than three  times the number  of
AGN  observed in  the Keck  fields.   We were   able to obtain  modest
agreement only  by taking the   lowest bounds on  the proximity effect
measurements and by including a high galactic escape fraction: $f_{\rm
esc} \gsim 20 \%$.  Conversely, if the flux decrement rate is adopted,
there is little room for a significant  stellar component, and $f_{\rm
esc} \lsim 5\%$ is required to match the faint AGN counts.  Future AGN
searches and   developments in our  ability  to measure the background
intensity  will be   useful   for  further constraining the    stellar
contribution to the ionizing background at high-$z$. 

We used our derived luminosity  functions and local determinations  of
the black hole  relic density to  determine the typical AGN efficiency
of converting mass  to light.  Although  many previous estimates based
on X-ray counts have obtained $\epsilon\gsim  0.1$ (e.g.  Elvis et al.
2002),   we find significantly  lower  values  $\epsilon\simeq 0.028 -
0.056$, suggesting that more than half of all  AGN are obscured in the
UV/optical.  Alternatively,  lower than expected values  of $\epsilon$
may  be derived if  much of a  black hole's mass is   set by a massive
``seed'' before  the onset of  its active phase.   A joint analysis of
UV-background  measurements, optical  counts,  and X-ray  observations
would likely resolve this issue. 

A similar analysis allows us to limit the fraction of black holes that
could have been active at any redshift.   For our models that are most
consistent with faint  AGN counts, we get $f_{\rm  on} \gsim 0.5\%$ at
$z\sim2.5$.  We can interpret  this in terms of a  lower limit  on the
average AGN  lifetime: $t_{agn} \gsim  10^7\rm{yrs}$.  For comparison,
an AGN luminosity   function    that matches the   proximity    effect
background (but over-produces the  faint counts) requires $f_{\rm  on}
\gsim 5\%$ and $t_{agn} \gsim  10^8\rm{yrs}$.  Note that these numbers
are firm  lower limits  because we assume  that there  are no obscured
AGN.   If a fraction  of AGN were obscured in  the optical, then these
limits would be increased by the inverse of the same fraction.

Finally, we conclude with some remarks about CDM-based theories of AGN
formation and how they  might  be constrained.  Usually   cosmological
models of this kind rely  on simple mappings  between dark halo masses
and AGN luminosities, but a fundamental unknown concerns the amount of
feedback that takes place (e.g. Haehnelt \&  Rees 1993, Haiman \& Loeb
1998).   We point  out  that one  potentially worrisome difficulty for
models with very  little feedback arises  from  direct observations of
the AGN LF at bright luminosities.  Specifically, the bright-end slope
measured by  the  SDSS at high  redshift   is even flatter  than  that
measured by the 2dF at low redshift.  This is  the opposite of what is
expected  if the AGN LF maps  simply to the   halo mass accretion rate
function (see,  e.g., Haiman  \&  Loeb 1998).  Therefore  the SDSS+2dF
results alone seem to suggest that  some feedback is needed for bright
to intermediate luminosity AGN. 

Another question is whether any  (additional) feedback might be needed
in low-mass/ low-luminosity objects in  order to explain the  observed
ionizing background rate at high-$z$.  As  discussed above, our Models
A and B span what we  believe to be  the range allowed by the ionizing
rate measurements.  Model B corresponds to the case where AGN dominate
the background, and, interestingly, its  evolution is very reminiscent
to that of  the halo mass function in  CDM: faint AGN are numerous  at
early times,  but  their numbers  fall-off  slowly at late times  (see
Figure 8).   It   would seem that   if  AGN do dominate  the  ionizing
background,  then CDM-models would  require  no additional feedback on
low mass  scales (relative to  what might already  be required  at the
high-mass end in order to match the SDSS results.)

If instead stars  contribute  with an escape fraction  consistent with
the Steidel et al.  (2002)  report (Model A),  then many fewer AGN are
permitted, and CDM-based  models would need  to be adjusted further in
order  to additionally suppress  the  formation of low-luminosity AGN.
Both Haiman, Madau  \& Loeb (1999)   and Kauffmann \&  Haehnelt (2000)
presented models with this  kind of luminosity-dependent feedback, but
it   is  unclear if, as   presented,  these models  would   be able to
reproduce both the ionizing rate (with an allowance for stars) and the
faint AGN counts in the  HDF and LBG  fields.  It would be interesting
to  see such a comparison.  The  model of Kauffmann \& Haehnelt (2000)
is  ideally   suited for  this   test because  it  predicts  the joint
population    of      galaxies  and  AGN      in   a   self-consistent
manner.\footnote{Indications that obscuration  plays   a key role   in
determining which AGN are  seen in the optical  add an extra  layer of
difficulty in this already difficult problem.} 

It is becoming increasingly clear that galaxy formation, AGN activity,
and  the resulting  ionizing  background are  intimately connected and
take part in an   important  feedback loop (Gebhardt  et   al.  2000a;
Ferrarese \&  Merritt   2000; Kauffmann  \&  Haehnelt 2000;   Bullock,
Kravtsov, \&  Weinberg 2000; Benson  et  al.  2002; Somerville  2002).
For this reason the need is high for self-consistent models that treat
all of these  processes   together, as  is  the need   for  additional
observational   constraints on  models    of  this kind.   The  limits
presented here  provide one small step in  this direction, but if some
agreement  can be reached  on measurements  of the ionizing background
emissivity, the resulting constraints would prove remarkably important
for piecing together the story of cosmological structure formation. 

\acknowledgments We would like to acknowledge helpful discussions with
Tom Abel,  Alberto Conti,  Piero  Madau, Smita  Mathur, Pat  McDonald,
Jordi  Miralda-Escud{\'    e},  Pat   Osmer,  Rick     Pogge, Marianne
Vestergaard, and Terry Walker.  We especially thank David Weinberg for
many  useful suggestions  that greatly  improved  the  quality of this
work.  We were supported by U.S.  DOE Contract No.DE-FG02-91ER40690.

%\end{references}

\begin{appendix}
%\section{Appendix:  IGM Absorption}

The following comes  primarily from MHR,  but see  also  Fardal et al.
(1998).   The  attenuation   due to photoelectric   absorption  can be
modelled by a distribution of discrete absorbers (or ``clouds''): 

\beq
  \tau_{eff}(\nu,z,\bar{z})= \int^{\bar{z}}_z  dz^\prime \int^\infty_0
     dN_{\rm{HI}} \frac{\partial^2 N}{\partial  N_{\rm{HI}}   \partial
     z^\prime} (1-e^{-\tau(\nu^\prime)}) 
\label{eq:tau_eff}
\eeq

\noindent 
where $\nu^\prime=\nu \      (1+z^\prime)/(1+z)$, and, for     a given
absorber, $N_{\rm{HI}}$  is  the column density  in  neutral hydrogen.
This distribution is usually parameterized as: 

\beq
  \frac{\partial^2   N}{\partial N_{\rm{HI}}  \partial      z^\prime}=
                   \frac{1}{N_o}
                   \left(\frac{N_{\rm{HI}}}{N_o}\right)^{-\beta}
                   (1+z^\prime)^{\gamma} 
\eeq 

\noindent 
If we assume that the distribution is smooth  over redshift and column
density, then  we get an    analytical expression for  the   effective
optical     depth.    We assume    $N_o=1.6\times10^{15}\rm{cm}^{-2}$,
$\beta=1.5$,  $\gamma=2$, which  MHR  claim  approximate more  precise
formulations.    [In \S \ref{sect:vary}   we compare  what our results
would be  using the corresponding  model from Fardal   et al.  (1998):
$N_o=3.27\times10^{14}\rm{cm}^{-2}$, $\beta=1.5$, $\gamma=2.58$.]   As
for  the attenuation  in a particular    cloud, we will  consider only
H\,{\sc i} and He\,{\sc ii} absorption: 

\beq
  \tau(\nu^\prime)=   N_{\rm{HI}}    \sigma_{\rm{HI}}(\nu^\prime)    +
                       N_{\rm{HeII}} \sigma_{\rm{HeII}}(\nu^\prime) 
\eeq

\noindent 
The cross  sections are  zero below  threshold ($\nu_{\rm{H}}=13.6$eV;
$\nu_{\rm{HeII}}=54.4$eV),  and above  threshold  are proportional  to
$\nu^{-3}$,                                                       with
$\sigma_{\rm{H}}(\nu_{\rm{H}})=6.35\times10^{-18}{\rm{cm}}^2$      and
$\sigma_{\rm{HeII}}(\nu_{\rm{HeII}})=1.59 \times 10^{-18}{\rm{cm}}^2$.
These cross sections are separated far enough in frequency that we can
consider  the ionization state of hydrogen  and helium separately.  We
can make the approximation that the relevant clouds are optically thin
(see Haardt \& Madau 1996; Fardal et al. 1998), so that: 

\beq
  \eta    \equiv        \frac{N_{\rm{HeII}}}{N_{\rm{HI}}}     =   0.45
       \frac{\Gamma_{\rm HI}}{\Gamma_{\rm HeII}} 
\eeq

\noindent 
This relation will not hold true in high column density clouds because
of  self-shielding  effects, but these  clouds  are  less abundant and
therefore  are less important  to the  effective  attenuation.  In any
case, a more careful study of the  helium ionization state should take
into account the self-shielding and reemission in the IGM. 

\end{appendix}

\end{document}